\documentclass[letterpaper]{article}
\usepackage{aaai2026}
\nocopyright

\usepackage{times}  \usepackage{helvet}  \usepackage{courier}  \usepackage[hyphens]{url}  \usepackage{graphicx}    \usepackage{natbib}  \usepackage{caption}     \usepackage{algorithm}
\usepackage{algorithmic}
\usepackage{newfloat}
\usepackage{listings}
\DeclareCaptionStyle{ruled}{labelfont=normalfont,labelsep=colon,strut=off} 
\floatstyle{ruled}
\newfloat{listing}{tb}{lst}{}
\floatname{listing}{Listing}

\usepackage{xcolor}

\usepackage{amsfonts}
\usepackage{amsmath}
\usepackage{soul}
\usepackage[table]{xcolor}
\usepackage{booktabs}
\usepackage{adjustbox}
\usepackage{multirow}
\usepackage{tikz}
\usetikzlibrary{arrows.meta, positioning, calc}
\definecolor{offwhite}{RGB}{250, 250, 240}

\newcommand*\emptydot[1][0.8ex]{\tikz\draw (0,0) circle (#1);} 

\newcommand*\fulldot[1][0.8ex]{\tikz\fill (0,0) circle (#1);}

\title{Control-Theoretic Content Moderation}
\author{
    Benedetta Tessa\textsuperscript{\rm 1, \rm 2},
    Serena Tardelli\textsuperscript{\rm 1},
    Marco Avvenuti\textsuperscript{\rm 2},
    Anna Monreale\textsuperscript{\rm 2},
    Stefano Cresci\textsuperscript{\rm 1}
}
\affiliations{
    \textsuperscript{\rm 1}IIT-CNR, Pisa, Italy\\
    \textsuperscript{\rm 2}University of Pisa, Italy\\
    benedetta.tessa@phd.unipi.it, serena.tardelli@iit.cnr.it, marco.avvenuti@unipi.it, anna.monreale@unipi.it, stefano.cresci@iit.cnr.it
    
}

\begin{document}
\maketitle

\begin{abstract}
A sizable literature studies content moderation locally, at the level of individual moderation decisions, for example by measuring or predicting the effects of specific interventions. However, the problem of how such decisions should be combined into effective platform-level moderation strategies is comparatively unexplored. We address this latter problem by formulating content moderation as a global, adaptive, and sequential decision process in which heterogeneous interventions must jointly balance multiple competing objectives. Drawing on feedback control, we introduce a general control-theoretic framework for composing moderation actions according to their expected effects on an evolving platform. We instantiate the framework in large-scale, empirically grounded simulations and compare two control-theoretic moderators against several baselines and local strategies. When moderation aims to maintain competing platform-level properties around desired conditions, the control-theoretic approaches achieve the best overall performance. They also use severe interventions more selectively and recover more effectively after external surges of harmfulness. These results demonstrate the advantages of studying content moderation as a global, adaptive, and sequential decision problem.
\end{abstract}

\section{Introduction}
\label{sec:intro}
Content moderation is a complex legal-socio-technical process encompassing multiple tasks: from detecting violating content and deciding if and how to intervene~\cite{wohn2017handle,kozyreva2023resolving}, to estimating the effects of those interventions~\cite{tessa2025beyond}, understanding how users respond~\cite{jhaver2023personalizing,cima2025investigating}, and assessing platform regulatory compliance~\cite{trujillo2026disarranged,kaushal2024automated}. These are essential matters to study, as they provide accurate ways to evaluate and optimize individual moderation decisions. Yet, they capture only part of the problem. For example, estimating the effects of a ban does not directly determine when or how extensively bans should be used at large, while comparing the effects of warnings and removals does not prescribe when one should be preferred over the other. In practice, platforms must continuously decide how to intervene across large volumes of content exhibiting different types and degrees of harm~\cite{gillespie2020content}, while user behavior, external events, and the effects of previous moderation decisions are themselves intertwined and evolving. As a consequence, locally effective decisions do not automatically compose into a \textit{globally} effective strategy~\cite{gao2016universal}. An open challenge is therefore to determine how heterogeneous moderation actions should be combined so that, collectively, they produce desirable outcomes at the platform level, within a complex and evolving environment.

This composition problem is inherently dynamic. However, while recent research increasingly considers temporal effects~\cite{costello2024durably,tardelli2024temporal}, it rarely formulates moderation as an \textit{adaptive} and sequential decision process. Most existing approaches are open-loop, as interventions are studied primarily based on the content, user, or threat being addressed, without systematically accounting for possible previous decisions or the state of the broader platform environment. In other words, moderation decisions do not incorporate feedback from the system they moderate. On the contrary, \textit{closed-loop} approaches offer a natural way to account for how interventions could alter subsequent behavior and how platform dynamics change in response to endogenous or external events~\cite{wang2020efficient}.

A further challenge concerns what moderation should ultimately optimize. Quantitative studies typically evaluate moderation against a narrow set of outcomes, such as reducing toxicity, misinformation, polarization, or improving the accuracy of enforcement~\cite{kumar2024watch,cima2025investigating,king2026beyond}. However, moderation simultaneously affects multiple outcomes and stakeholders, whose interests, values, and requirements differ and often conflict. For instance, recent research shows that curbing toxic behavior may also affect harmless participation and legitimate expression~\cite{trujillo2022make,cima2025investigating}, while platforms must additionally contend with economic incentives and regulatory constraints~\cite{gorwa2024politics}. Prior theoretical work has documented such tensions and the inherently \textit{multi-objective} nature of moderation~\cite{jiang2023trade}. Yet, we lack quantitative frameworks for jointly representing \textit{competing objectives} and studying how moderation decisions should balance, and ultimately act upon, them.

\textbf{Research Focus.} These limitations have a common root. Quantitative content moderation has predominantly been studied at a microscopic, single-decision level~\cite{furutani2026impact}, rather than as a platform level decision problem. Here we shift the level of abstraction and formulate content moderation macroscopically, as the feedback control of an evolving legal-socio-technical system. Control theory provides a natural foundation for this formulation, as it studies how interventions can regulate the behavior of dynamic systems toward desired conditions by repeatedly observing their state and acting upon the resulting feedback~\cite{astrom2008feedback}. We thus introduce a general computational framework for control-theoretic content moderation that is simultaneously:
\begin{itemize}
\item \textbf{global}---composing individual interventions based on their collective effects on the platform state;
\item \textbf{adaptive} and \textbf{closed-loop}---deciding interventions over time, based on feedback from the moderated system;
\item \textbf{multi-objective}---jointly modeling and accounting for multiple competing objectives.
\end{itemize}
We instantiate this framework in a set of simulations grounded in empirical estimates from prior moderation research, where control-theoretic moderators repeatedly select among heterogeneous interventions according to their expected consequences. Across different evolutions of the platform environment, we compare this adaptive approach with alternative moderation strategies and examine their aggregate performance, their intervention choices, and the evolution of platform-level outcomes over time.

\textbf{Contributions.} We make the following contributions:
\begin{itemize}
    \item \textbf{Problem formulation.} We formulate content moderation as a global, closed-loop, and multi-objective sequential decision problem, shifting the focus from optimizing individual moderation decisions to composing them into adaptive platform-level strategies.
    \item \textbf{Control-theoretic framework.} We introduce a general and extensible framework for control-theoretic content moderation that operationalizes this formulation. The framework accommodates heterogeneous interventions, competing objectives, probabilistic effects, proportionality, and feedback-driven adaptation over time.
    \item \textbf{Empirical evaluation.} We instantiate the framework in a set of empirically grounded simulations and evaluate it under different platform dynamics, comparing control-theoretic moderation against alternative strategies.\footnote{The code will be made publicly available upon acceptance.} Control-theoretic moderators achieve the lowest overall losses across two experimental scenarios, selectively composing heterogeneous interventions to balance competing objectives and enabling substantially better platform recovery after an exogenous shock.
\end{itemize}
 \section{Related Work}
\label{sec:relatedwork}

\subsection{Simulations of Moderation Strategies}
A few studies used control theory as part of content moderation simulations. We discuss these studies first, as they are most related to our present work, before turning to other computational approaches. \citet{wang2020efficient} modeled the spread of positive and negative information and used feedback to dynamically allocate interventions. They focused on the overarching goal of containing negative information at minimum expense, and applied interventions based on predefined rules. Others formulated recommendation as a control problem. \citet{sprenger2024control} compared a model-free strategy that responds to current user opinions with a Model Predictive Control (MPC) strategy that also considers their future evolution, with recommendations optimized for engagement. Building on this model, \citet{pagan2026misinformation} additionally considered misinformation mitigation and its trade-off with engagement. These studies motivate the use of feedback control for specific combinations of threats and interventions, but address narrow instances of the content moderation problem. Our work generalizes this perspective into a global, dynamic, and multi-objective content moderation framework.

Beyond control-theoretic approaches, agent-based and LLM simulations have examined how predefined moderation strategies shape online behavior and platform outcomes over time~\cite{fidone2026evaluating,pollacci2026simulating,truong2026audit}. By modeling the downstream effects of interventions on subsequent outputs, interactions, and content exposure, these approaches capture consequences that static evaluations overlook. However, their moderation rules remain fixed or locally triggered, rather than adapting to feedback from the overall platform state or balancing competing objectives.
Other agent-based and network simulations examined how local interventions shape broader patterns of information diffusion across communities and platforms~\cite{murdock2024agent,furutani2026impact}. These simulations examined how communities and information diffusion change under moderation, but did not use those changes as feedback to inform subsequent decisions.

\subsection{Broader Content Moderation Studies}
Beyond simulation studies, a vast literature investigated how real moderation interventions affected users and online communities.
These studies show that moderation can reduce harmful behavior, but may also alter participation or produce unintended consequences~\cite{cima2024great}.
Moreover, the effects of the same intervention can vary substantially across users and over time~\cite{gleason2025suspense}.

More specifically, some studies have examined how moderation affects subsequent behavior, content visibility, and participation.
For example, warning-labeled posts received more engagement than unlabeled posts~\cite{zannettou2021won}.
Visibility reduction can lower engagement with content while leaving it available~\cite{thero2022investigating},
while content removals and the explanations accompanying them are associated with subsequent user behavior and perceptions of fairness~\cite{jhaver2019does,jhaver2019did}.
Community restrictions and bans expose further trade-offs.
Reddit quarantines, restrictions, and bans can reduce harm or participation, but may also displace behavior elsewhere~\cite{chandrasekharan2022quarantined,horta2021platform}.
Studies of deplatforming reveal further heterogeneity, as many users leave or reduce toxicity after deplatforming~\cite{chandrasekharan2017you,jhaver2021evaluating}, while smaller groups may temporarily increase their toxic behavior~\cite{thomas2023disrupting,cima2025investigating}.
No action is therefore uniformly effective across users, outcomes, and contexts, supporting gradual and personalized moderation~\cite{cresci2022personalized,jhaver2023personalizing}.

These findings inform our simulations, where moderation actions have distinct effects on harmfulness, activity, and engagement. 
Our framework translates estimates of individual effects into platform-level decisions about when and how broadly to intervene and how to combine actions. \section{Control-Theoretic Moderation}
\label{sec:control-theory}

\subsection{Problem Definition}
Our goal is to develop a computational framework for studying how individual moderation decisions can be combined into effective platform-level strategies. To this end, we consider an abstract moderator as the entity making these decisions. The moderator may represent a human decision-maker, an automated system, or a combination of the two. The moderator manages a platform where a large population of users produces content, some of which may be harmful and become moderation targets. For each target, the moderator must decide whether to intervene and, if so, which intervention to apply among the available ones. These decisions are made repeatedly as new content is produced and the platform evolves, with the aim of simultaneously maintaining multiple platform-level properties around their desired values. Examples of such properties are the prevalence of harmful content or the extent of user activity on the platform. Deciding which combination of interventions to apply is nontrivial, because interventions differ in their effects and may simultaneously improve some properties while worsening others~\cite{trujillo2022make,cima2025investigating}. Consequently, na\"ive strategies such as intervening on every target or always applying the strongest intervention, would not produce desirable platform-level outcomes.

We formalize this problem in discrete time, with $t=1,\ldots,T$. Let $\mathbf{s}_t$ denote the platform properties of interest at time $t$, and $\mathbf{r}$ their desired values. At each $t$, let $\mathcal{M}_t$ denote the set of moderation targets and $\mathcal{A}$ the set of interventions available, including taking no action. The moderator selects an intervention $a_m\in\mathcal{A}$ for each target $m \in \mathcal{M}_t$. The set of interventions selected at $t$ is denoted by $\mathbf{a}_t$. Interventions may affect one or more platform properties, thereby influencing $\mathbf{s}_t$. We seek a moderation policy $\mathbf{a}^*$ that selects these interventions so as to minimize, over time, the deviation of the platform properties from their desired values:
\begin{equation}
\label{eq:problem-definition}
    \mathbf{a}^*
    =
    \arg\min_{\mathbf{a}}
    \sum_{t=1}^{T}
    \mathcal{L}\!\left(\mathbf{s}_t,\mathbf{r}\right),
\end{equation}
where $\mathcal{L}$ is a loss function that quantifies deviations from the desired values across the multiple considered objectives. The problem is therefore not to identify the best intervention in isolation, but their best composition over time to maintain desired platform-level conditions as the platform evolves.

\subsection{Content Moderation as Feedback Control}
This problem resembles classical feedback control. The analogy provides a natural mapping between content moderation and control theory, a well-established mathematical and engineering framework for regulating the behavior of dynamical systems through interventions and feedback~\cite{astrom2008feedback}. In control-theoretic language, the platform and its users represent the \textit{controlled system}, and their joint evolution determines the platform-level \textit{state} observed by the moderator. The desired values define the \textit{reference}, while moderation interventions constitute the available \textit{control actions}. By affecting the platform and its users, such actions influence how the state evolves. Finally, the moderator acts as the \textit{controller}, selecting actions based on the observed state and the moderation objectives.

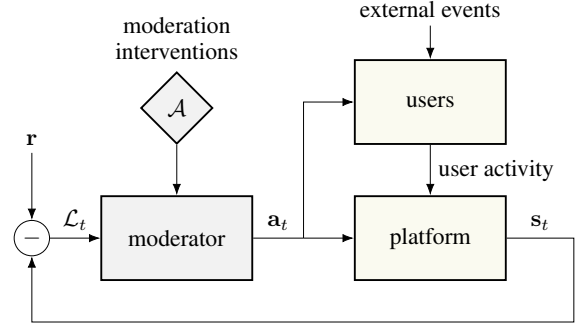
\begin{figure}[t]
\centering
    \adjustbox{max width=.9\columnwidth}{\tikzset{
  block/.style={
    draw,
    thick,
    rectangle,
    minimum height=1.5em,
    minimum width=2em,
    align=center
  },
  sum/.style={
    draw,
    circle,
    minimum size=1.5em,
    inner sep=0pt
  },
  input/.style={coordinate},
  output/.style={coordinate},
}

\begin{tikzpicture}[auto, node distance=2.5cm, >=Latex]

\node[sum] (sum) {\( - \)};
  \node[above=1cm of sum] (reference) {\( \mathbf{r} \)};

\node[
    block,
    right=0.75cm of sum,
    minimum width=2.25cm,
    minimum height=1.25cm,
    fill=gray!10
  ] (controller) {moderator};

\node[
    block,
    right=1.5cm of controller,
    minimum width=2.25cm,
    minimum height=1.25cm,
    fill=offwhite
  ] (platform) {platform};

\node[
    block,
    above=0.75cm of platform,
    minimum width=2.25cm,
    minimum height=1.25cm,
    fill=offwhite
  ] (users) {users};

\node[
    above=0.5cm of users
  ] (external) {external events};

\node[
    block,
    above=1.3cm of controller,
    minimum size=0.75cm,
    rotate=45,
    inner sep=0pt,
    anchor=center,
    fill=gray!10
  ] (actions) {\rotatebox{-45}{\( \mathcal{A} \)}};

\draw[->] (reference) -- (sum);
  \draw[->] (sum) -- node[above] {\( \mathcal{L}_t \)} (controller);

\draw[->]
    (actions)
    node[above=0.6cm, align=center] {moderation\\interventions}
    -- (controller.north);

\coordinate[right=0.75cm of controller] (actionbranch);

  \draw[-]
    (controller.east)
    -- node[above] {\( \mathbf{a}_t \)}
    (actionbranch);

  \draw[->]
    (actionbranch)
    |- (users.west);

  \draw[->]
    (actionbranch)
    -- (platform.west);

\draw[->]
    (external)
    -- (users);

\draw[->]
    (users)
    -- node[right, align=left] {user activity}
    (platform);

\draw[->]
    (platform.east)
    -- ++(1,0)
    node[midway, above] {\( \mathbf{s}_t \)}
    |- ++(0,-1.25)
    -| (sum.south);

\end{tikzpicture}
 }
    \caption{\textbf{Control-theoretic content moderation.} The moderator observes the current state of the platform $\mathbf{s}_t$ and compares it with the reference $\mathbf{r}$ to select among the available interventions $\mathcal{A}$. Interventions can affect the platform and its users, whose behavior is also influenced by external events. User activity contributes to the subsequent evolution of the platform, whose updated state is observed again to inform future moderation decisions, closing the feedback loop.}
    \label{fig:moderation-loop}
\end{figure}

The defining feature of this formulation is feedback. As represented in Figure~\ref{fig:moderation-loop}, at each time step the moderator observes the current platform state, evaluates it against the reference, and selects moderation actions accordingly. These actions are applied to the controlled system, which continues to evolve under the combined influence of moderation, user behavior, and external events. The resulting state is then observed again and used to inform the next moderation decision, thus closing the loop. Consequently, the moderator need not determine a fixed sequence of interventions in advance, based on predefined rules. Instead, it repeatedly makes decisions according to how the platform actually evolves. This distinction is central to online content moderation, where intervention effects are uncertain, user behavior may change abruptly, and unexpected events may require adapting moderation strategies~\cite{gillespie2020content}.

\subsection{A Control-Theoretic Moderation Framework}
Having defined the problem and established its connection to feedback control, we now formalize the components of a general control-theoretic moderation framework.

\textbf{States, objectives, and interventions.}
We represent the platform at time $t$ through a state vector $\mathbf{s}_t\in\mathbb{R}^n$, whose dimensions correspond to the $n$ platform-level properties that moderation seeks to regulate (e.g., harmful content prevalence, user activity, etc.). Each user $i$ is similarly associated with a user state $\mathbf{u}_{i,t}$ describing properties relevant to their behavior (e.g., their tendency to post or produce harmful content). The desired platform conditions are specified by a reference $\mathbf{r}\in\mathbb{R}^n$. Weights in $\mathbf{w}\in\mathbb{R}_{\geq 0}^n$ assign a priority to each objective, allowing to account for multiple objectives simultaneously without assuming that they are equally important. At every time step, the set $\mathcal{M}_t$ contains the available moderation targets. For each target $m\in\mathcal{M}_t$, the moderator selects one action $a_m$ from the intervention set $\mathcal{A}$. 

\textbf{System dynamics.}
Moderation actions in $\mathcal{A}$ may affect the platform, its users, both, or neither. For example, a content removal directly alters the platform, while a warning message acts on the targeted user and may influence their subsequent behavior. Instead, taking no action on a target produces neither effect. We represent these dynamics as
\begin{align}
    \mathbf{u}_{i,t+1} &= f_u(\mathbf{u}_{i,t},\mathbf{a}_t,
    \boldsymbol{\epsilon}_{i,t}), \notag\\
    \mathbf{s}_{t+1} &= f_s(\mathbf{s}_t,\mathbf{U}_{t+1},
    \mathbf{a}_t,\boldsymbol{\xi}_t),
\label{eq:system-dynamics}
\end{align}
where $\mathbf{U}_t$ collects the user states and $\boldsymbol{\epsilon}_{i,t}$ and $\boldsymbol{\xi}_t$ are error terms that capture unmodeled or stochastic variation. To select actions, the moderator uses a model of these dynamics to estimate the consequences of candidate interventions, yielding predicted future states $\hat{\mathbf{s}}_{t+h}$. Such a model may be informed by prior empirical evidence or learned from observations~\cite{cerulli2026theory}, and is not assumed to accurately capture the actual system dynamics. Thus, in general, $\hat{\mathbf{s}}_{t+h} \neq \mathbf{s}_{t+h}$, reflecting uncertainty in the effects of moderation interventions and in the evolution of the platform. Actions are selected based on how close their predicted outcomes bring the platform to the reference. This is quantified via a weighted multi-objective loss
\begin{equation*}
    \mathcal{L}_{\mathrm{obj}}(\hat{\mathbf{s}},\mathbf{r})
    =
    \sum_{j=1}^{n}
    w_j(\hat{s}_j-r_j)^2.
\end{equation*}

\textbf{Proportionality.}
An appropriate moderation strategy should consider not only the expected effects of an intervention, but also whether the intervention is \textit{proportionate} to the severity of the violation. Overly severe responses to minor violations may be undesirable even when they improve platform-level outcomes. This principle is well established in theoretical literature~\cite{kiesler2012regulating,marique2020sanctions}. Here we operationalize it by incorporating proportionality directly into the moderator's objective. Let $v_m$ denote the severity of moderation target $m$ (e.g., the toxicity score of a post). For each available intervention $a \in \mathcal{A}$, we define $v_a^{\min}$ and $v_a^{\max}$ as the minimum and maximum violation severity for which $a$ is considered proportionate. We then define the proportionality penalty as
\begin{equation*}
P(a,v_m) =
\begin{cases}
0,
    & \text{if } v_a^{\min} \leq v_m \leq v_a^{\max}, \\[2mm]
(v_a^{\min}-v_m)^2,
    & \text{if } v_m < v_a^{\min}, \\[2mm]
(v_m-v_a^{\max})^2,
    & \text{if } v_m > v_a^{\max}.
\end{cases}
\end{equation*}
$P(a,v_m) = 0$ when the intervention is proportionate to the violation and increases quadratically with the distance from the nearest boundary otherwise. We incorporate this penalty into the loss as
\begin{equation}
\label{eq:loss}
\begin{aligned}
    \mathcal{L}_{\mathrm{prop}}
    &=
    \sum_{m\in\mathcal{M}_t}
    P(a_m,v_m),
    \\
    \mathcal{L}_t
    &=
    \mathcal{L}_{\mathrm{obj}}(\hat{\mathbf{s}}_t,\mathbf{r})
    +
    w_{\mathrm{prop}}\,\mathcal{L}_{\mathrm{prop}}.
\end{aligned}
\end{equation}
where $\mathcal{L}_{\mathrm{prop}}$ is the proportionality loss and $w_{\mathrm{prop}}\geq0$ controls its importance relative to the platform-level objectives. 

Proportionality is modeled as a loss term rather than a hard optimization constraint, allowing the moderator to occasionally select a disproportionate intervention when its expected contribution to the other objectives outweighs the corresponding penalty. Further, the proportionality ranges of different interventions may overlap, allowing multiple interventions to be considered proportionate for the same target.

\textbf{Planning over time.}
Moderation interventions can have effects that persist beyond the time at which they are applied. Straightforwardly, banning a user precludes their future participation. Similarly, other interventions also alter subsequent user activity~\cite{trujillo2022make}. Selecting actions based solely on their immediate effects may therefore overlook long-term consequences. To account for these effects and operationalize the general optimization problem defined in Eq.~\eqref{eq:problem-definition}, the moderator solves it repeatedly over a finite lookahead horizon $H$, rather than optimizing a single time step or the entire platform trajectory at once. At time $t$, it predicts the platform trajectory until $t+H$. This prediction accounts for the platform evolution and the planned sequence of interventions at each time step. This sequence is selected so as to minimize the cumulative loss over $H$
\begin{equation}
\label{eq:fine-horizon-optimization}
    \mathbf{A}^{*}_t
    =
    \arg\min_{\mathbf{A}_{t:t+H-1}}
    \sum_{h=1}^{H}
    \mathcal{L}_{t+h}, \end{equation}
where $\mathbf{A}_{t:t+H-1}$ denotes a sequence of moderation actions over the next $H$ time steps. However, while planning a sequence of interventions over $H$ time steps, only the actions selected for the current time step $t$ are actually applied. The remaining actions are reevaluated at $t+1$, when the moderator observes the new platform state, updates its predictions, and solves the optimization problem over $H$ again. This is an instance of Model Predictive Control (MPC), a classical control approach that repeatedly optimizes actions over a finite receding horizon~\cite{rawlings2017model}. It combines lookahead planning with feedback, as decisions account for their expected long-term consequences while being continuously revised according to how the platform actually evolves.

\subsection{Design Principles, Modularity, and Extensibility}
This framework enables moderation systems that are \textit{global}, by jointly reasoning over a multidimensional platform state and multiple moderation targets; \textit{dynamic} and \textit{closed-loop}, by repeatedly adapting interventions based on feedback from the evolving system; and \textit{multi-objective}, by jointly accounting for multiple desired platform properties and moderation objectives. This formulation defines the core components of the framework rather than a specific implementation. Its modular design allows these components to be instantiated or extended in different ways, depending on the moderation setting and research questions. Appendix~\ref{sec:appendix-extensions} formalizes several such extensions. Accordingly, implementation-specific choices---e.g., which platform and user properties are modeled, which interventions are available and how their effects are represented, how future states are predicted---are deliberately left open here and specified in the next section, when instantiating the framework.
 \section{Simulation and Experimental Settings}
\label{sec:settings}
Here we provide the high-level experimental settings of the simulation. Appendix~\ref{sec:appendix-calibration} reports further details.

\subsection{Simulation Environment}
\label{sec:settings-simulation}

\textbf{Simulation design.}
We simulate the evolution of an online platform populated by a fixed initial set of $N=10{,}000$ users over $T=100$ time steps. Each experiment is repeated over $r=10$ independent runs. The random seed is fixed for all moderators within each run, but varied across runs. Hence, on each run all moderators operate on independent copies of the same initial user population and the same underlying stochastic realization of user behavior. In other words, absent moderation, users would produce the same posts across moderators. Their realized behavior will however diverge due to the applied moderation actions. This design allows differences within a run to be attributed to moderation while capturing stochastic variability across runs.

\textbf{Platform properties.}
We model three platform-level properties: content \textit{harmfulness}, user \textit{activity}, and user \textit{engagement}. These dimensions capture distinct and competing interests in platform governance. Reducing harmful content contributes to platform safety, maintaining activity preserves user participation and platforms' economic interests, maintaining engagement captures the social response that users receive from peers~\cite{jiang2023trade}. These properties are modeled to be interdependent: moderation interventions that reduce harmfulness may also suppress activity~\cite{trujillo2022make}, while the engagement received by a post may depend on its harmfulness~\cite{avalle2024persistent}. Therefore they provide a compact setting in which moderators must balance competing objectives. We initially present results obtained with objectives that all have equal weight $w=1$, including proportionality ($w_\mathrm{prop}=1$), while Appendix~\ref{app:sensitivity-objective-weights} reports sensitivity analyses with different weights. 

\textbf{Reference values.}
For each platform property, we define a reference value representing the state that moderators seek to maintain. We set the activity and engagement references to their average values in the absence of moderation, estimated through preliminary unmoderated simulations. In contrast, we set a considerably stricter harmfulness reference, well below its unmoderated level and corresponding to an average post harmfulness of $\bar{v}=0.3$. These reference values deliberately create a demanding multi-objective task: moderators must reduce harmfulness while preserving the activity and engagement of the unmoderated platform. Appendix~\ref{app:sensitivity-reference-values} reports results of sensitivity analyses that consider alternative reference values.

\textbf{User properties.}
Each user is characterized by its propensity to be active, harmful, and to receive engagement on the platform. The activity propensity is sampled from a Beta distribution and determines the probability that the user produces a post at any given time step. The harmfulness of a user's post is sampled from a user-specific log-normal distribution clipped to $[0,1]$. Following~\citet{avalle2024persistent}, engagement is determined by a user-specific parabolic function of post harmfulness, allowing the relationship between harmfulness and engagement to be nonlinear while also capturing heterogeneity in the engagement different users tend to receive. This modeling of user properties produces heterogeneous behavior, with a bulk of low propensity users and relatively high propensities concentrated among smaller user groups. In the absence of moderation, the user-specific parameters remain fixed throughout the simulation. Their distributions and functional forms are calibrated from empirical findings, as detailed in Appendix~\ref{app:simulation-environment}.

\begin{table}[t]
\centering
\small
\setlength{\tabcolsep}{5pt}
\begin{tabular}{lccccc}
\toprule
& \multicolumn{2}{c}{\textbf{severity}}  & \textbf{platform state}  & \multicolumn{2}{c}{\textbf{user behavior}} \\
\cmidrule(lr){2-3} \cmidrule(lr){4-4} \cmidrule(l){5-6}
\textbf{action $a_m$} & $v_a^{\min}$ & $v_a^{\max}$ & $s_t^{(h,a,e)}$ & $u_{i,t+1}^{(h)}$ & $u_{i,t+1}^{(a)}$ \\
\midrule
no action & 0.0 & 0.2 & 0        & 0        & 0        \\
warning   & 0.2 & 0.4 & 0        & $-10\%$  & 0        \\
demotion  & 0.4 & 0.6 & $-75\%$  & 0        & 0        \\
removal   & 0.6 & 0.8 & $-100\%$ & $-20\%$  & $-20\%$  \\
ban       & 0.8 & 1.0 & $-100\%$ & 0        & $-100\%$ \\
\bottomrule
\end{tabular}
\caption{Platform and user effects of choosing moderation action $a_m$ to moderate target post $m$ by user $u_i$. The severity interval $[v_a^{\min},v_a^{\max}]$ denotes the range of post harmfulness for which action $a_m$ is considered proportionate. Platform state effects refer to $m$'s contribution to the platform's harmfulness ($h$), activity ($a$), and engagement ($e$). User effects are relative changes in $u_i$'s future harmfulness and activity.}
\label{tab:interventions-effects}
\end{table}
 
\textbf{Moderation interventions.}
For each post, moderators can select among five actions: \textit{no action}, \textit{warning}, \textit{demotion}, post \textit{removal}, and user \textit{ban}. Actions may affect the current platform state, the moderated user's future behavior, both, or neither. As summarized in Table~\ref{tab:interventions-effects}, taking no action leaves both unchanged. A warning does not affect the current platform's state, but slightly reduces the user's future harmfulness~\cite{yildirim2023short}. A demotion greatly reduces a post's visibility on the platform, affecting its contribution to the overall platform's harmfulness, activity, and engagement, but does not affect subsequent user behavior~\cite{thero2022investigating}. Removal eliminates the current post and reduces the user's future harmfulness and activity~\cite{jhaver2019does,jhaver2019did}. Finally, a ban removes the post and permanently prevents any future activity from the user.

The effects of these actions are calibrated from recent empirical measurements on real platforms, as reported in Appendix~\ref{app:simulation-environment}. Effects on future user behavior are probabilistic, and Table~\ref{tab:interventions-effects} reports their means. Consequently, individual user responses may deviate from the reported mean effects and, with low probability, an intervention may even backfire, as in reality~\cite{cima2025investigating}. For simplicity, we assumed independent, additive, and permanent effects (i.e., intervention effects do not decay over time).

\subsection{Moderators}
\label{sec:settings-moderators}
We compare six moderators representing progressively richer approaches to composing individual moderation interventions into platform-level strategies. Our goal is not to identify an optimal moderator, but to examine how different formulations of the content moderation problem, summarized by their main capabilities in Table~\ref{tab:main-results}, translate into system-level outcomes. We therefore consider strategies ranging from local rules that are agnostic to their aggregate consequences, to feedback controllers that explicitly reason over the platform state and the expected effects of alternative interventions. Further implementation details of the following moderators are provided in Appendix~\ref{app:control-theorethic-mod}.

\textbf{No moderation.}
Baseline where the moderator never intervenes. It provides the counterfactual evolution of the platform in the absence of moderation.

\textbf{Fixed removal.}
The moderator removes every post whose harmfulness exceeds a fixed threshold $\tau$ and takes no action otherwise. Following recent works about online toxicity~\cite{nogara2025toxic}, we set $\tau=0.7$. This represents a simple content-level enforcement rule in which decisions depend exclusively on whether individual posts exceed a predefined harmfulness threshold.

\textbf{Proportional.}
The moderator selects actions exclusively according to post harmfulness, applying increasingly severe actions to increasingly harmful content. Given a moderated post having harmfulness $v_m$, it selects the action $a$ from Table~\ref{tab:interventions-effects} such that $v_a^{\min} \le v_m < v_a^{\max}$. This represents a richer local strategy in which actions are always proportionate to individual violations, but their aggregate platform-level consequences are not considered.

\textbf{Adaptive removal.}
The moderator retains removal as its only action but dynamically adjusts the harmfulness threshold according to the deviation of platform harmfulness from its reference. When harmfulness exceeds its reference, the threshold $\tau$ decreases and removal becomes more aggressive. When it falls below its reference, $\tau$ increases. This isolates the value of platform-level feedback while retaining a single action and objective.

\textbf{MPC-based.}
As a predictive control-theoretic implementation of our framework, the MPC-based moderator selects actions by considering both their immediate and expected future consequences. For each post, it evaluates every available action according to its immediate effect on the platform state, its predicted effect on subsequent user behavior and platform state, its proportionality, and an estimate of its longer-term cost over horizon $H=50$. It then applies the action with the lowest estimated loss. The longer-term cost is estimated by comparing the expected future loss following each action with that under no intervention and scaling this difference over $H$, providing a computationally lightweight approximation of Eq.~\eqref{eq:fine-horizon-optimization}.

\textbf{PID-based.}
As an alternative control-theoretic implementation of our framework, we consider a proportional-integral-derivative (PID) controller, one of the most widely used feedback-control approaches~\cite{astrom2008feedback}. At each time step, the PID-based moderator combines the current deviation from the desired platform state (\textit{proportional}), its accumulated deviation over time (\textit{integral}), and its change since the previous time step (\textit{derivative}) to produce a continuous corrective signal. Because moderation actions are discrete and heterogeneous, we translate this signal into interventions by evaluating candidate actions according to their proportionality and expected effects over horizon $H=50$. This provides an alternative instantiation of the framework in which moderation responds not only to the current platform state, but also to the persistence and evolution of previous deviations. Appendix~\ref{app:sensitivity-time-horizon} reports results of varying the size of the lookahead horizon $H$.

\subsection{Experimental Scenarios}
\label{sec:settings-scenarios}
We comparatively evaluate moderators under two scenarios that differ in the dynamics of the platform environment.

\textbf{Stationary platform.}
User behavior follows the stochastic processes dictated by their properties, without additional external perturbations. Platform harmfulness, activity, and engagement fluctuate over time as users probabilistically post and generate heterogeneous content, but their underlying distributions remain stationary. This scenario represents ordinary platform operation, where moderators must maintain the desired state despite endogenous variability.

\textbf{Surge of harmfulness.}
We also consider a dynamic environment in which users experience a temporary exogenous increase in their propensity to produce harmful content. The perturbation increases smoothly, peaks approximately halfway through the simulation, and subsequently subsides, producing substantially larger variation in platform harmfulness than under stationary conditions. This scenario captures periods in which external events temporarily alter online behavior, such as contentious elections or breaking events that trigger heightened conflict and harmful discourse. 

\begin{table*}[t]
\centering
\small
\setlength{\tabcolsep}{5pt}
\begin{tabular}{lccccrrc*{4}{c}}
\toprule
& \multicolumn{4}{c}{\textbf{capabilities}$^\dagger$} & & \multicolumn{6}{c}{\textbf{evaluation measures} \textit{(lower is better)}} \\
\cmidrule{2-5} \cmidrule{7-12}
\textbf{moderator} & FB & MA & MO & PL & \textbf{actions} & harm. & act. & eng. & \cellcolor{gray!15} total & prop. & \cellcolor{gray!15} total + prop. \\
\midrule
\textit{Stationary platform} &  &  &  &  &  &  &  &  & \cellcolor{gray!15} &  & \cellcolor{gray!15} \\ [0.5ex]
No moderation       & -- & -- & -- & -- & 0       & .462 & \textbf{.000} & \textbf{.000} & \cellcolor{gray!15}.462 [.457, .467] & .158 & \cellcolor{gray!15}.621 [.615, .627] \\
Fixed removal       & \emptydot & \emptydot & \emptydot & \emptydot & 40{,}882 & \textbf{.000} & .340 & .305 & \cellcolor{gray!15}.645 [.638, .652] & .032 & \cellcolor{gray!15}.677 [.670, .684] \\
Proportional        & \emptydot & \fulldot & \emptydot & \emptydot & 46{,}819 & \textbf{.000} & .154 & .161 & \cellcolor{gray!15}.315 [.309, .321] & \textbf{.000} & \cellcolor{gray!15}.315 [.309, .321] \\
Adaptive removal    & \fulldot & \emptydot & \emptydot & \emptydot & 37{,}835 & \textbf{.000} & .312 & .276 & \cellcolor{gray!15}.589 [.584, .594] & .037 & \cellcolor{gray!15}.625 [.620, .631] \\
MPC-based           & \fulldot & \fulldot & \fulldot & \fulldot & 71{,}566 & \textbf{.000} & .038 & .042 & \cellcolor{gray!15}\underline{.080 [.078, .081]} & \underline{$>.00$} & \cellcolor{gray!15}\underline{.080 [.078, .082]} \\
PID-based           & \fulldot & \fulldot & \fulldot & \fulldot & 55{,}516 & \underline{.007} & \underline{.025} & \underline{.029} & \cellcolor{gray!15}\textbf{.061 [.058, .065]} & .004 & \cellcolor{gray!15}\textbf{.065 [.062, .068]} \\
\midrule
\textit{Surge of harmfulness} &  &  &  &  &  &  &  &  & \cellcolor{gray!15} &  & \cellcolor{gray!15} \\ [0.5ex]
No moderation       & -- & -- & -- & -- & 0       & 1.296 & \textbf{.000} & \textbf{.028} & \cellcolor{gray!15}1.324 [1.319, 1.329] & .242 & \cellcolor{gray!15}1.565 [1.560, 1.571] \\
Fixed removal       & \emptydot & \emptydot & \emptydot & \emptydot & 54{,}778  & \textbf{.000} & .640 & .623 & \cellcolor{gray!15}1.263 [1.261, 1.265] & .035 & \cellcolor{gray!15}1.298 [1.296, 1.300] \\
Proportional        & \emptydot & \fulldot & \emptydot & \emptydot & 60{,}548  & \textbf{.000} & .622 & .619 & \cellcolor{gray!15}1.241 [1.239, 1.243] & \textbf{.000} & \cellcolor{gray!15}1.241 [1.239, 1.243] \\
Adaptive removal    & \fulldot & \emptydot & \emptydot & \emptydot & 54{,}154  & \textbf{.000} & .642 & .625 & \cellcolor{gray!15}1.267 [1.265, 1.269] & .043 & \cellcolor{gray!15}1.310 [1.308, 1.312] \\
MPC-based           & \fulldot & \fulldot & \fulldot & \fulldot & 116{,}312 & \textbf{.000} & .453 & .455 & \cellcolor{gray!15}\underline{0.908 [0.905, 0.911]} & \underline{.004} & \cellcolor{gray!15}\underline{0.912 [0.909, 0.915]} \\
PID-based           & \fulldot & \fulldot & \fulldot & \fulldot & 112{,}256 & \underline{.016} & \underline{.308} & \underline{.325} & \cellcolor{gray!15}\textbf{0.649 [0.580, 0.717]} & .012 & \cellcolor{gray!15}\textbf{0.661 [0.594, 0.728]} \\
\bottomrule
\multicolumn{12}{l}{$^\dagger$ FB: feedback, MA: multiple moderation actions, MO: multiple objectives, PL: planning}
\end{tabular}
\caption{\textbf{Comparison of moderation strategies across the two experimental scenarios}: stationary platform and surge of harmfulness. The left-hand side summarizes the main capabilities of each moderator, while the right-hand size reports quantitative results aggregated over $r=10$ runs. For each moderator, we report the mean number of moderation actions applied per run and its ability to steer the platform toward the desired reference values, measured through the resulting losses (lower values are better). Losses are reported both for the individual components and in aggregate, with the total loss shown both excluding and including proportionality. Quantitative results are means across runs. For the two total losses we additionally report 95\% confidence intervals. For each scenario and loss measure, the best result is shown in \textbf{bold} and the second best is \underline{underlined}.}
\label{tab:main-results}
\end{table*}
 
\begin{figure}[t]
    \centering
    \includegraphics[width=.9\columnwidth]{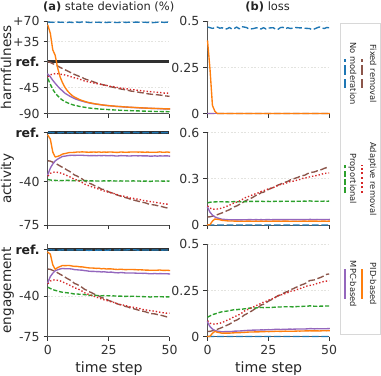}
    \caption{\textbf{Evolution of platform states and losses}, for harmfulness, activity, and engagement. Stationary platform. \textbf{(a)} Deviations from reference values and \textbf{(b)} corresponding losses. To highlight differences, only the first 50 time steps are shown. Values are averaged across simulation runs.}
    \label{fig:states-losses-stationary}
\end{figure}

\subsection{Evaluation}
\label{sec:settings-evaluation}
We evaluate moderators both in terms of the platform outcomes they produce and the interventions through which they achieve them. For platform outcomes, we report the losses associated with harmfulness, activity, engagement, and proportionality separately, as well as the resulting total loss. For each experiment we report mean values and $95\%$ confidence intervals across runs. We also perform pairwise comparisons between moderators in each experimental scenario, testing differences in total loss using two-sided paired $t$-tests with Holm correction for multiple comparisons.
Appendix~\ref{sec:appendix-sensitivity} reports additional sensitivity analyses.
 \section{Results}
\label{sec:results}
We evaluate how effectively the moderators steer the platform toward the predefined objectives under stationary conditions and an external surge of harmfulness. Table~\ref{tab:main-results} summarizes the overall results across both scenarios.

\subsection{Stationary Platform}
\textbf{Platform outcomes.}
The stationary scenario exposes the trade-offs inherent in moderation. Without moderation, activity and engagement remain at their reference values, but harmfulness incurs a substantial loss (.462). Conversely, Fixed Removal, Proportional, and Adaptive Removal meet the harmfulness reference, but at the cost of large deviations in activity and engagement. The trajectories in Figure~\ref{fig:states-losses-stationary} further show that MPC and PID rapidly reduce the initial harmfulness deviation while avoiding the progressive deterioration of activity and engagement observed with removal-based strategies. The control-theoretic moderators thus achieve a substantially better balance. PID obtains the lowest total loss (.061), followed by MPC (.080), compared to .315 for the next-best strategy. Notably, PID accepts a small harmfulness loss (.007) while substantially reducing activity (.025) and engagement (.029) losses. This demonstrates that allowing limited deviations on one objective can yield substantially better outcomes overall. MPC instead eliminates harmfulness and proportionality losses while accepting slightly larger activity and engagement losses, showing that different moderators can realize different trade-offs within the same problem formulation. Finally, Table~\ref{tab:main-results} shows that performance does not depend on intervention volume alone. MPC applies the most actions, while PID achieves lower total loss with fewer, indicating that platform outcomes depend not simply on how much moderation is applied, but on how interventions are selected and composed.

\begin{figure}[t]
    \centering
    \includegraphics[width=\columnwidth]{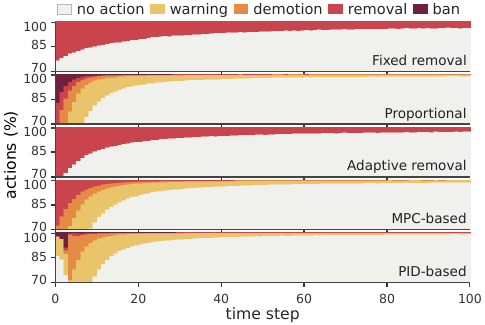}
    \caption{\textbf{Distribution of moderation actions over time.} Stationary platform. Stacked areas show the relative proportions of actions applied at each time step. The \textit{y}~axis is truncated at 70\% to better highlight differences. The omitted portion consists mainly of no-action decisions. Values are averaged across simulation runs.}
    \label{fig:actions-time-stationary}
\end{figure}

\begin{figure}[t]
    \centering
    \includegraphics[width=\columnwidth]{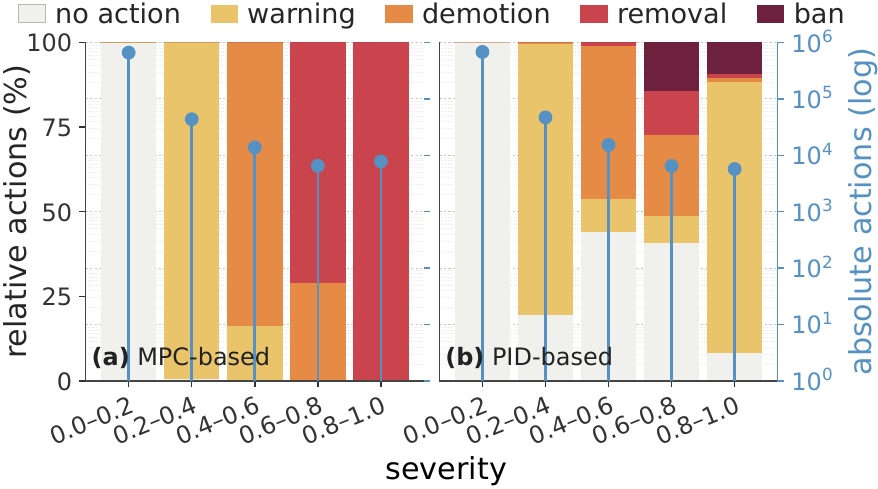}
    \caption{\textbf{Distribution of moderation actions by severity of the moderated post.} Stationary platform. In each severity bin, stacked bars show relative proportions, while teal lollipops show absolute action counts. \textbf{(a)} MPC-based and \textbf{(b)} PID-based moderators. Values are averaged across runs.}
    \label{fig:actions-severity-stationary}
\end{figure}

\textbf{Moderation behavior.}
Figure~\ref{fig:actions-time-stationary} shows that all moderators intervene strongly early in the simulation and progressively reduce moderation, as the persistent effects of early interventions reduce the need for subsequent actions. The main difference between moderators therefore lies not in this temporal pattern, but in which interventions are selected. Compared with Proportional, MPC and PID favor milder actions. MPC never uses bans, while PID uses them only briefly at the beginning. This more selective composition of interventions explains their lower overall losses, as stronger actions are not necessary to reach the desired platform conditions and can impose larger costs on other objectives. PID also exhibits a brief cold-start period, as its integral and derivative components initially lack sufficient error history, but rapidly stabilizes thereafter. Figure~\ref{fig:actions-severity-stationary} further shows that both controllers generally increase intervention severity with post severity, although MPC follows the proportionality ranges more closely, while PID selects a broader mixture of actions across severity levels. Accordingly, as per Table~\ref{tab:main-results}, MPC incurs essentially no proportionality loss, whereas PID accepts a minor loss there, while achieving the lowest total loss.

\begin{figure}[t]
    \centering
    \includegraphics[width=.9\columnwidth]{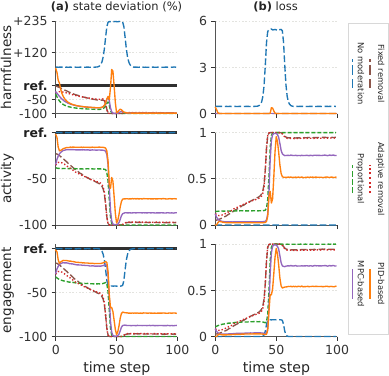}
    \caption{\textbf{Evolution of platform states and losses}, for harmfulness, activity, and engagement. Surge of harmfulness. \textbf{(a)} Deviations from reference values and \textbf{(b)} corresponding losses. Values are averaged across simulation runs. }
    \label{fig:states-losses-surge}
\end{figure}

\begin{figure}[t]
    \centering
    \includegraphics[width=\columnwidth]{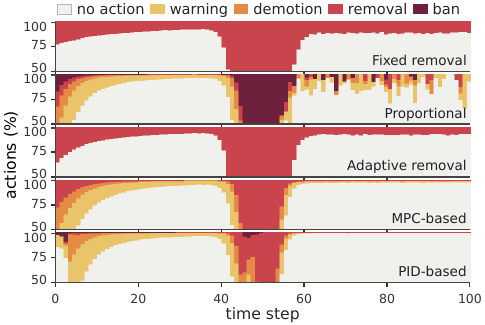}
    \caption{\textbf{Distribution of moderation actions over time.} Surge of harmfulness. Stacked areas show the relative proportions of actions applied at each time step. The \textit{y}~axis is truncated at 50\% to better highlight differences. The omitted portion consists mainly of no-action and post removal decisions. Values are averaged across simulation runs.}
    \label{fig:actions-time-surge}
\end{figure}

\begin{figure}[t]
    \centering
    \includegraphics[width=\columnwidth]{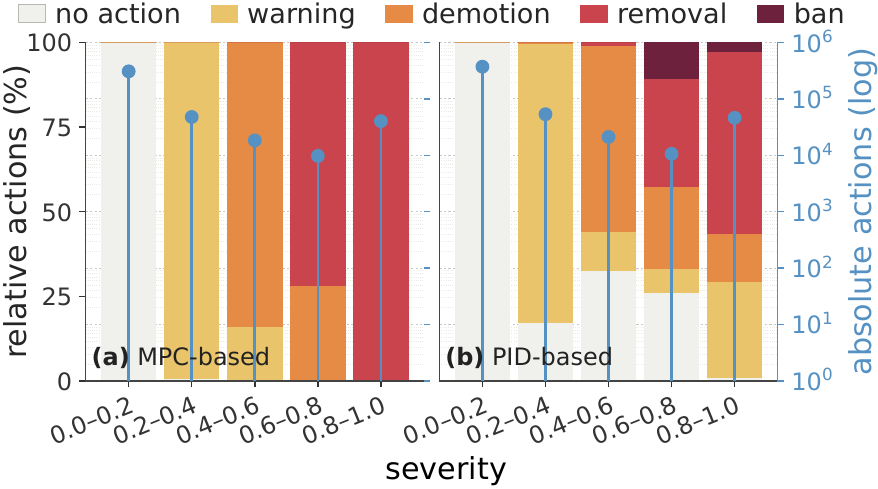}
    \caption{\textbf{Distribution of moderation actions by severity of the moderated post.} Surge of harmfulness. In each severity bin, stacked bars show relative proportions, while teal lollipops show absolute action counts. \textbf{(a)} MPC-based and \textbf{(b)} PID-based moderators. Values are averaged across runs.}
    \label{fig:actions-severity-surge}
\end{figure}

\subsection{Surge of Harmfulness}
\textbf{Platform outcomes.}
The surge of harmfulness creates a substantially more challenging scenario, with all moderators incurring higher total losses than under stationary conditions, as reported in Table~\ref{tab:main-results}. The perturbation is clearly visible in Figure~\ref{fig:states-losses-surge}, for the unmoderated platform trajectory. Despite this disruption, the control-theoretic moderators retain a clear advantage. PID achieves again the lowest total loss (.649), followed by MPC (.908), compared with 1.241 for the next-best strategy. Their trajectories further reveal important differences in recovery. While most moderation strategies suppress the harmfulness surge, they sharply reduce activity and engagement and largely fail to restore them afterward. PID instead recovers both properties substantially better than the other moderators, accepting a small harmfulness loss (.016) in exchange for markedly lower activity and engagement losses. Thus, suppressing harmfulness alone does not imply successful platform regulation: effective moderation must also account for the state in which the platform is left after the perturbation.

\textbf{Moderation behavior.}
Figure~\ref{fig:actions-time-surge} shows that the harmfulness surge triggers a sharp increase in moderation across all moderators. Notably, this response is not specific to adaptive moderators, as even fixed and local strategies react because the surge increases the number and severity of moderation targets. What differs is \textit{how} moderators respond. Proportional mechanically escalates toward bans as more posts enter the highest severity ranges, whereas MPC and PID respond predominantly through removals. Thus, the advantage of adaptive moderation lies not in reacting to the perturbation per se, but in composing that response differently, which helps explain the different post-surge recovery trajectories observed in Figure~\ref{fig:states-losses-surge}. Figure~\ref{fig:actions-severity-surge} further shows how intervention choices changed relative to the stationary scenario. MPC retains nearly the same action distribution across severity levels, responding to the surge primarily by increasing intervention volume. PID instead changes its action composition, especially for more severe posts ($0.6$--$0.8$ and $0.8$--$1.0$), for which it shifts toward stronger interventions. This provides quantitative evidence that the appropriate intervention depends not only on the local characteristics of the moderation target, but also on the broader state of the platform.

\subsection{Sensitivity Analyses}
Additional sensitivity analyses reported in Appendix~\ref{sec:appendix-sensitivity} show that these results are largely robust to alternative parameterizations, while also illustrating how design choices shape moderation behavior. MPC and PID remain relatively stable across different harmfulness reference values, although PID is more sensitive to changes in objective weighting. Increasing the proportionality weight progressively aligns interventions with their prescribed severity ranges, demonstrating how objective configuration can systematically alter moderation behavior. Finally, introducing a prediction horizon ($H>0$) substantially changes intervention choices compared to no lookahead ($H=0$), while longer horizons produce comparatively limited additional changes.
 \section{Discussion and Conclusions}
\label{sec:discussion}

\textbf{Scope and contribution.}
Through extensive and empirically grounded simulations, we showed the value of casting content moderation as the problem of composing individual decisions into effective platform-level strategies over time. To this end, our main contribution is a computational framework for studying this composition problem quantitatively. This direction of research opens new questions, including how heterogeneous interventions interact, how their aggregate consequences unfold, and whether multiple---potentially conflicting---outcomes can be maintained simultaneously~\cite{jiang2023trade}. However, our work does not prescribe an optimal moderation strategy or identify the best combination of interventions. Nor does it determine what platforms or society ought to optimize. Desirable platform conditions, competing priorities, legitimate interventions, and acceptable constraints remain inherently normative and contextual~\cite{gillespie2020content,cima2025contextualized}. Still, the framework makes these choices explicit and operational, providing a quantitative means to study the strategies, trade-offs, and consequences that follow from them.

\textbf{Connecting different strands of research.}
This perspective is particularly timely as growing empirical research characterizes individual components of moderation, from identifying problematic content to estimating intervention effects and user responses. Our framework complements these efforts by providing a system-level model in which their findings can be combined. Computational social science, HCI/CSCW, and ML/AI can inform platform states, user dynamics, and intervention effects~\cite{king2026beyond,jhaver2019does,tessa2025beyond}. Governance research can inform objectives and constraints~\cite{gillespie2020content,jiang2023trade}. Control theory provides mature computational methods for composing these elements into adaptive moderation strategies~\cite{astrom2008feedback,rawlings2017model}. As knowledge about individual moderation decisions accumulates, understanding how these decisions compose at the platform level becomes an increasingly important step.

\textbf{Beyond control-theoretic moderation.}
We formulated content moderation as a dynamic platform-level decision problem and explored control theory as one possible approach to solving it. Our formalization and results suggest that it represents a suitable approach to this problem, since its core components map directly onto those of platform-level moderation. However, other computational paradigms could also support adaptive moderation. For example, reinforcement learning could learn intervention policies through repeated interactions~\cite{farajtabar2017fake}, while agent-based models could represent user interactions and network dynamics within which adaptive moderation strategies can be studied~\cite{murdock2024agent,fidone2026evaluating}. Thus, control theory is but one of multiple promising approaches to study moderation as a dynamic system-level problem in which individual interventions are composed and adapted as the environment evolves.

\textbf{Multi-objective moderation.}
A further implication of our formulation is that there is no context-independent notion of ``better moderation'' without specifying which outcomes matter and how competing interests should be balanced~\cite{trujillo2022make}. While these tensions are increasingly recognized in theoretical and governance research~\cite{gillespie2020content,jiang2023trade}, quantitative approaches still tend to evaluate moderation against comparatively narrow outcomes~\cite{king2026beyond}. We argue that quantitative moderation research should more explicitly account for multiple---and especially \textit{competing}---objectives. Moderation interventions rarely affect a single outcome in isolation, while platforms must simultaneously navigate commercial incentives, operational constraints, and regulatory obligations~\cite{gorwa2024politics}. Evaluating interventions only along few dimensions can thus obscure trade-offs and provide incomplete accounts of their system-level effects. Multi-objective formulations instead make these trade-offs objects of quantitative analysis, allowing to study the practical implications of different priorities.

\textbf{Limitations and future work.}
Our implementation of the proposed framework necessarily simplifies the full complexity of real platforms. For example, our simulations consider a limited set of properties and interventions, rely on empirically informed but simplified dynamics and specific parameter configurations, and make several assumptions about intervention effects (e.g., their independence, additivity, and persistence). They also abstract away social interactions, and are therefore not intended to study phenomena such as polarization or information diffusion. However, these limitations concern our specific implementation rather than the problem formulation or the general framework. Different or richer models, including network and agent-based models, could be used within the same framework, broadening the scope of control-theoretic moderation studies. The proposed framework is indeed modular and extensible. Appendix~\ref{sec:appendix-extensions} formalizes several possible extensions, including intervention costs and budgets, moderation delays, and incomplete knowledge of intervention effects.

These possibilities suggest several directions for future research. More realistic and empirically grounded models could broaden the moderation phenomena that can be studied and provide computational laboratories for investigating strategies and trade-offs that are difficult to examine on real platforms. Further, beyond evaluating existing interventions, the framework could inform the design of new ones. Exploring hypothetical interventions could reveal when the available action space is insufficient and what properties new interventions would need. Finally, models calibrated with observations from real communities could help compare plausible mechanisms or moderation strategies according to their ability to reproduce observed platform dynamics, using simulation not only to explore possible futures but also to better understand observed ones.

\section{Acknowledgments}
This work is partly supported by the European Union with the ERC project DEDUCE under grant \#101113826.

\bibliography{references}


\appendix
\setcounter{secnumdepth}{2}

\section{Framework Extensions}
\label{sec:appendix-extensions}
The core control-theoretic framework can be extended to accommodate additional characteristics of real-world moderation and relax some underlying assumptions. Here we formalize three such extensions: limited moderation resources, delayed moderation, and incomplete knowledge of intervention effects. These extensions illustrate how additional operational constraints and sources of uncertainty can be incorporated into the framework. While neither of these extensions have been implemented or experimentally evaluated in this work, they nonetheless contribute to demonstrating the extensibility and generality of the framework.

\subsection{Intervention Costs and Moderation Budget}
\label{app:budget}
Moderation interventions may differ substantially in the resources required to apply them, making operational cost an important consideration in moderation at scale~\cite{gillespie2020content}. Cost can be modeled simply in terms of the number of interventions performed, or more generally by assigning different costs to different interventions. For example, some interventions may require human review or additional operational effort, while others may be applied at comparatively low cost. A moderator operating under limited resources could therefore account not only for the expected outcomes of candidate interventions, but also for their costs.

Let $c_a \geq 0$ denote the cost of intervention $a$. The total cost of the interventions selected at time $t$ is
\begin{equation*}
    C_t
    =
    \sum_{m\in\mathcal{M}_t} c_{a_m}.
\end{equation*}
Given an available moderation budget $B_t$, limited resources can be modeled as a hard constraint on the optimization problem defined in Eq.~\eqref{eq:fine-horizon-optimization}:
\begin{equation*}
\begin{aligned}
    \mathbf{A}^{*}_t
    &=
    \arg\min_{\mathbf{A}_{t:t+H-1}}
    \sum_{h=1}^{H}\mathcal{L}_{t+h},
    \\
    \text{s.t.}\qquad
    &\sum_{h=0}^{H-1} C_{t+h} \leq B_t.
\end{aligned}
\end{equation*}
This formulation is appropriate when the available budget represents a strict resource limit that cannot be exceeded. Alternatively, intervention cost can itself be treated as an optimization objective captured by a loss term. Defining $\mathcal{L}_{\mathrm{cost},t}=C_t$, the general loss in Eq.~\eqref{eq:loss} can be extended as
\begin{equation*}
    \mathcal{L}_t
    =
    \mathcal{L}_{\mathrm{obj}}(\hat{\mathbf{s}}_t,\mathbf{r})
    +
    w_{\mathrm{prop}}\,\mathcal{L}_{\mathrm{prop}}
    +
    w_{\mathrm{cost}}\,\mathcal{L}_{\mathrm{cost},t},
\end{equation*}
where $w_{\mathrm{cost}}\geq0$ determines the weight assigned to reducing moderation costs. This formulation does not impose a fixed spending limit, but instead allows intervention costs to be traded off against the other moderation objectives. The two approaches can also be combined, allowing the moderator to minimize costs while remaining within a hard resource constraint.

\subsection{Delayed Moderation}
\label{app:delays}
We assumed that a moderation decision immediately affects the controlled system. In practice, several forms of delay may occur. A violation may require time to be reviewed before a decision is made, an intervention may be enforced only after some delay, or its consequences may become observable only after a certain time has passed~\cite{trujillo2025transparency,trujillo2022make}. Although these mechanisms are operationally distinct, they can be accounted for within the same framework by modeling the time between the identification of a moderation target and the moment at which the corresponding intervention begins affecting the system. Such delays make moderation more challenging, as actions must anticipate future platform conditions rather than only respond to the current state.

Let $d_a \geq 0$ denote the delay associated with intervention $a$, measured in time steps. If $a$ is selected for a target at time $t$, its effect enters the system dynamics starting at time $t+d_a$. Then, the predicted system state can be expressed by modifying Eq.~\eqref{eq:system-dynamics} as
\begin{equation*}
    \hat{\mathbf{s}}_{t+h}
    =
    \hat{f}_s
    \left(
        \hat{\mathbf{s}}_{t+h-1},
        \hat{\mathbf{U}}_{t+h},
        \mathbf{a}^{\,\mathrm{eff}}_{t+h-1}
    \right),
\end{equation*}
where $\mathbf{a}^{\,\mathrm{eff}}_{t}$ denotes the moderation effects that are active at time $t$, including effects of interventions selected at earlier time steps whose delays have elapsed. The optimization itself retains the same form of Eq.~\eqref{eq:fine-horizon-optimization}, but future states, and consequently future losses, are now predicted according to the delayed realization of intervention effects. The lookahead horizon $H$ becomes particularly important in this setting. An intervention whose delay exceeds the remaining prediction horizon cannot contribute to the predicted platform outcomes within that horizon. Modeling delays therefore allows the moderator to distinguish between interventions according not only to their expected effects, but also to when those effects are expected to materialize.

\subsection{Learning Intervention Effects}
\label{app:learning-effects}
To solve the optimization problem defined in Eq.~\eqref{eq:fine-horizon-optimization}, the moderator must estimate how candidate interventions will affect future platform states. In our experiments, intervention effects are probabilistic rather than deterministic, mirroring real-world uncertainty. However, their probability distributions are assumed to be known by the moderator. This assumption is partly motivated by the empirical literature on the effects of moderation interventions~\cite{jhaver2019did,jhaver2019does,thero2022investigating,trujillo2022make,yildirim2023short,cima2025investigating}, but remains a simplification. In real settings, intervention effects may be only partially anticipated~\cite{tessa2025beyond} and may themselves need to be learned.

This assumption can be relaxed by representing the system dynamics through unknown parameters $\boldsymbol{\theta}$. Rather than having access to their true values, the moderator maintains an estimate $\hat{\boldsymbol{\theta}}_t$ based on the effects observed up to time $t$. State predictions then become
\begin{align*}
    \hat{\mathbf{u}}_{i,t+h}
    &=
    \hat{f}_u
    \left(
        \hat{\mathbf{u}}_{i,t+h-1},
        \mathbf{a}_{t+h-1};
        \hat{\boldsymbol{\theta}}_t
    \right),
    \\
    \hat{\mathbf{s}}_{t+h}
    &=
    \hat{f}_s
    \left(
        \hat{\mathbf{s}}_{t+h-1},
        \hat{\mathbf{U}}_{t+h},
        \mathbf{a}_{t+h-1};
        \hat{\boldsymbol{\theta}}_t
    \right),
\end{align*}
where $\hat{\boldsymbol{\theta}}_t$ represents the moderator's current knowledge about intervention effects. After applying an intervention and observing the resulting system evolution, this knowledge can be updated:
\begin{equation*}
    \hat{\boldsymbol{\theta}}_{t+1}
    =
    \mathcal{U}
    \left(
        \hat{\boldsymbol{\theta}}_t,
        \mathbf{s}_t,
        \mathbf{U}_t,
        \mathbf{a}_t,
        \mathbf{s}_{t+1},
        \mathbf{U}_{t+1}
    \right),
\end{equation*}
where $\mathcal{U}$ denotes a generic learning or updating procedure.

Incomplete knowledge introduces an exploration--exploitation trade-off. Interventions may be selected not only because their currently estimated effects are expected to improve the platform objectives (exploitation), but also because observing their consequences can reduce uncertainty about intervention effects and improve future decisions (exploration). This can be incorporated through established approaches from adaptive control or reinforcement learning~\cite{mesbah2018stochastic}, for example by explicitly rewarding information acquisition or by optimizing decisions under uncertainty over the intervention-effect parameters. The framework does not prescribe how intervention effects should be learned. Depending on the setting, $\hat{\boldsymbol{\theta}}_t$ could be estimated from accumulated platform observations, experiments, prior empirical evidence, or combinations thereof, while uncertainty could be represented through point estimates or probability distributions. This extension transforms the moderator from a controller that acts on fixed knowledge of intervention effects into one that jointly learns those effects and exploits the acquired knowledge to regulate the platform.

\section{Simulation Calibration and Grounding}
\label{sec:appendix-calibration}

\subsection{Simulation Environment}
\label{app:simulation-environment}
This section provides further implementation and calibration details on our simulations.

\textbf{Platform properties.}
Each post $j$ has a harmfulness score $v_j\in[0,1]$ that resembles widely-used toxicity scores, such as those provided by Perspective API, Detoxify and other available tools~\cite{cima2024great,nogara2025toxic}. At each time step $t$, platform harmfulness $s_t^{(h)}$ is the sum of the harmfulness scores of all posts shared at $t$, activity $s_t^{(a)}$ is the total number of such posts, and engagement $s_t^{(e)}$ is their total received engagement. The platform state is consequently
\[
    \mathbf{s}_t =
    \left(
        s_t^{(h)},
        s_t^{(a)},
        s_t^{(e)}
    \right).
\]

\textbf{Loss normalization.}
The three platform properties are measured on different scales. Consequently, their raw deviations from the corresponding reference values can have markedly different magnitudes. Directly aggregating squared deviations would therefore make the resulting loss components difficult to compare, as a property measured on a larger scale could dominate the overall loss simply because of its numerical magnitude, rather than because its deviation from the desired state is relatively more important.

To place the three objectives on a comparable scale, we normalize each deviation by the corresponding reference value before computing its loss. Let $r^{(p)}$ denote the desired reference value for platform property $p$. We define
\[
L_p(s_t,r)
=
\left(
\frac{s_t^{(p)}-r^{(p)}}{r^{(p)}}
\right)^2,
\qquad
p \in \{h,a,e\}.
\]
The quantity inside the square therefore represents the relative deviation of property $p$ from its reference value. For example, a $10\%$ deviation from the harmfulness reference and a $10\%$ deviation from the activity reference have the same normalized magnitude, regardless of the absolute scales on which harmfulness and activity are measured. Squaring this relative deviation yields a dimensionless loss, allowing harmfulness, activity, and engagement to contribute comparably to the aggregate objective. For the same comparability reason, the proportionality loss is normalized by the number of interventions applied. This normalization makes the loss components numerically comparable. Then, their associated weights can be used separately to normatively prioritize or discount particular objectives relative to the others.

\begin{figure}[t]
    \centering
    \includegraphics[width=.85\columnwidth]{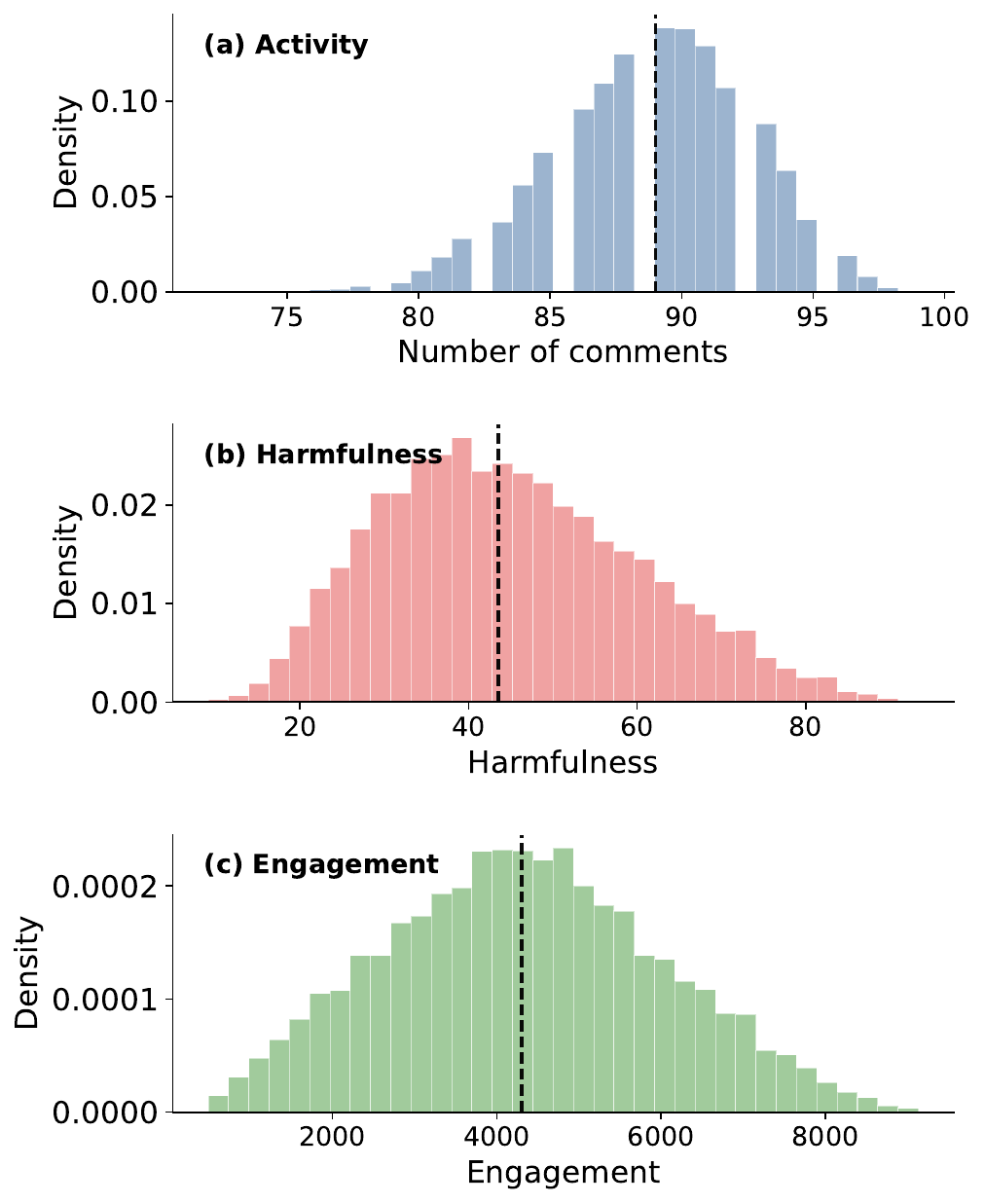}
    \caption{\textbf{User distributions.} Distribution of activity \textbf{(a)}, harmfulness \textbf{(b)}, and engagement \textbf{(c)} for an unmoderated group of users over a single simulation run.}
    \label{fig:user-distr}
\end{figure}

\textbf{User properties.}
Each user $u_i$ is characterized by three behavioral components governing their activity, harmfulness, and engagement. To model heterogeneity in activity across users, each user is assigned an individual activity propensity $p_i$, sampled once at initialization as
\[
p_i \sim \mathrm{Beta}(\alpha_a,\beta_a).
\]
This propensity remains fixed in the absence of moderation and represents the user's probability of posting at any given time step. Specifically, at each time $t$, we independently sample
\[
z_{i,t} \sim \mathrm{Bernoulli}(p_i),
\]
where $z_{i,t}=1$ indicates that user $i$ produces a post and $z_{i,t}=0$ otherwise. Thus, the Beta distribution determines heterogeneity in baseline activity across users, while the Bernoulli distribution realizes each user's posting behavior over time.

When $z_{i,t}=1$, the harmfulness of the resulting post is represented by a continuous score $h_{i,t} \in [0,1]$, resembling widely used toxicity scores such as those produced by Perspective API and Detoxify~\cite{nogara2025toxic}. To reproduce the strongly right-skewed, long-tailed distributions of toxicity observed empirically on online platforms, where most content exhibits low harmfulness while comparatively few posts reach high values~\cite{cima2025investigating}, we sample post harmfulness from a user-specific log-normal distribution:
\[
X_{i,t}\sim\mathrm{LogNormal}(\mu_i,\sigma_i),
\]
and bound the resulting value to $[0,1]$. The user-specific parameters $(\mu_i,\sigma_i)$ introduce heterogeneity in harmfulness across users, while repeated sampling from each user's distribution allows the harmfulness of their individual posts to vary over time.

Finally, the engagement obtained by a post is determined by a user-specific parabolic function of post harmfulness,
\[
e_{i,t}
=
a_i h_{i,t}^2 + b_i h_{i,t} + c_i,
\]
such that engagement increases with post harmfulness up to a user-specific peak, after which the relationship reverses and further increases in harmfulness result in progressively lower engagement~\cite{avalle2024persistent}. The coefficients $(a_i,b_i,c_i)$ vary across users, allowing heterogeneity in the engagement users tend to receive. Figure~\ref{fig:user-distr} illustrates the resulting distributions of activity, harmfulness, and engagement for a population of $N=10{,}000$ users over a single unmoderated simulation run.

\begin{figure}[t]
    \centering
    \includegraphics[width=\linewidth]{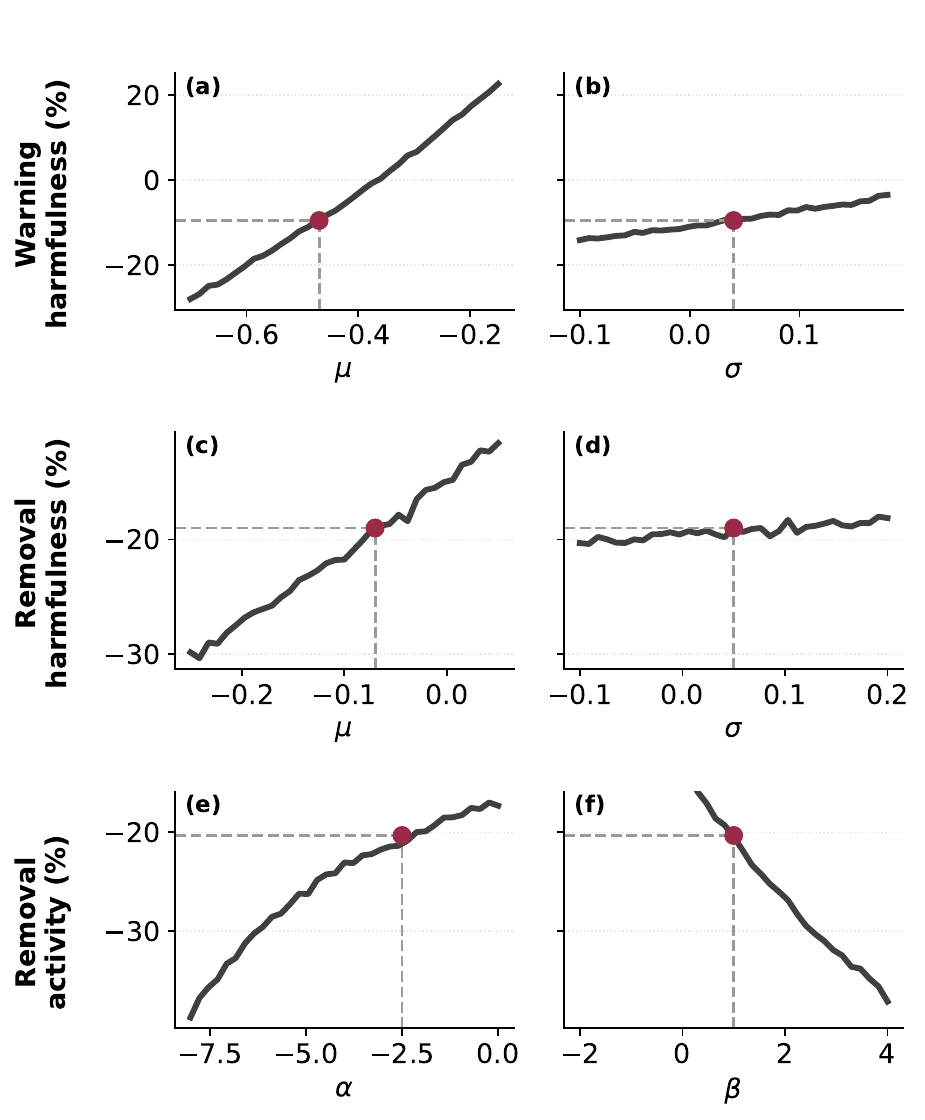}
    \caption{\textbf{Calibration of the user effects of warnings and removals.} Each panel relates the mean intervention-induced modification of one user parameter (\textit{x} axis) to the resulting average change in observable harmfulness or activity across Monte Carlo simulations (\textit{y} axis). The red markers indicate the selected values on the calibration curves, while dashed projection lines highlight the corresponding mean parameter change (\textit{x} axis) and the resulting mean effect (\textit{y} axis).}
\label{fig:intervention_sensitivity}
\end{figure}

\textbf{Calibrating moderation effects on users.}
Moderation interventions can affect both the current platform state and future user behavior. Effects on the current platform state can be applied directly. For example, removing a post eliminates that post's contribution to current activity, harmfulness, and engagement. Instead, effects on future user behavior require a different treatment. In our simulation, observable behaviors such as activity and harmfulness are stochastic outcomes generated from the parameters and distributions characterizing each user. An intervention therefore does not directly modify a user's realized future activity or harmfulness. Rather, it modifies the underlying parameters governing the corresponding user behavioral process, which in turn changes the distribution from which future behavior is generated.

To ground these modifications empirically, we draw on prior studies measuring how moderation interventions affect subsequent user behavior~\cite{yildirim2023short,thero2022investigating,jhaver2019did,jhaver2019does}. These studies provide the empirical targets reported as user effects in Table~\ref{tab:interventions-effects}, expressed as average changes in observable outcomes such as user activity and harmfulness. However, these empirical effect sizes do not directly specify how the underlying parameters of our generative model should change. We therefore need to determine which parameter modifications reproduce the desired behavioral effects.

We perform this calibration through Monte Carlo simulation. Intervention-induced changes to user parameters are themselves stochastic. Rather than applying the same parameter modification to every moderated user, we sample its magnitude from a normal distribution. Thus, users receiving the same intervention need not undergo identical parameter changes. For each intervention effect, we systematically vary the mean of this distribution and repeatedly simulate its application across the heterogeneous user population. We then measure the resulting average change in the corresponding observable outcome. This produces a mapping between the expected magnitude of the underlying parameter modification and its expected behavioral effect, from which we select the values that most closely reproduce the empirical targets in Table~\ref{tab:interventions-effects}. For example, interventions affecting activity modify the parameter $\alpha_a$ governing users' activity propensities. The intervention-specific change in $\alpha_a$ is sampled probabilistically around a calibrated mean. Changing $\alpha_a$ modifies $p_i$, which in turn changes users' probabilities of posting and, ultimately, aggregate activity.

Figure~\ref{fig:intervention_sensitivity} illustrates representative calibration curves. Each curve relates the mean intervention-induced modification of the relevant user parameter to the resulting average change in observable activity or harmfulness across Monte Carlo simulations. The selected mean parameter values are identified directly on the calibration curves. In this way, the calibration maps empirical estimates of the average behavioral effects of moderation interventions onto stochastic modifications of the parameters governing user behavior in our simulation.

\begin{algorithm}[t]
\caption{Simulation step at time $t$}
\label{alg:simulation-step}
\begin{algorithmic}[1]

\STATE Let $U_t$ be the user population at time $t$
\STATE Let $A$ be the set of available interventions
\STATE Initialize the set of moderation targets $M_t \leftarrow \emptyset$

\FOR{each non-banned user $u_i \in U_t$}
    \STATE Sample whether $u_i$ posts according to its activity propensity

    \IF{$u_i$ posts}
        \STATE Generate a moderation target $m$
        \STATE Sample its harmfulness $h_m$
        \STATE Compute its engagement $e_m$
        \STATE $M_t \leftarrow M_t \cup \{m\}$
    \ENDIF
\ENDFOR

\STATE Compute the platform state $s_t$ from the generated targets in $M_t$

\FOR{each target $m \in M_t$}
    \STATE Select intervention $a_m \in A$
    \STATE Apply $a_m$
\ENDFOR

\STATE Update $U_{t+1}$ to the resulting user population
\STATE Proceed to time $t+1$ and repeat the simulation step

\end{algorithmic}
\end{algorithm}

\textbf{Simulation step.}
Having defined how users generate content and how moderation interventions affect the platform and its users, we can now describe a complete simulation step. At each time $t$, every eligible user $u_i$ independently produces a post with probability $p_i$. For each generated post, harmfulness $h_{i,t}$ is sampled from the corresponding user's harmfulness distribution and engagement $e_{i,t}$ is computed from their engagement function. Aggregating the resulting posts yields the unmoderated platform state $\mathbf{s}_t$.

The moderator then processes each post $m \in \mathcal{M}_t$ and selects an intervention $a_m$. As described above, interventions can affect the current platform state, future user behavior, both, or neither. Platform-level effects are applied directly to $\mathbf{s}_t$ and compose additively across interventions, reflecting our assumption that their effects are independent and additive. User-level effects are instead implemented by modifying the behavioral parameters of the moderated user $u_i$, thereby changing the stochastic process governing their subsequent behavior. The resulting user states are carried forward into $\mathbf{U}_{t+1}$ and used to generate behavior at the next time step. Intervention-induced changes to user parameters persist in subsequent steps unless altered by later moderation, reflecting our assumption that user-level intervention effects are permanent. This completes the feedback loop through which current moderation decisions affect both the current platform state and its future evolution. Algorithm~\ref{alg:simulation-step} summarizes the complete process.
    
\subsection{Moderators}
\label{app:control-theorethic-mod}
This section provides additional technical implementation details for some of the moderators.

\textbf{Adaptive removal.}
The adaptive removal moderator dynamically adjusts the harmfulness threshold used for content removal according to the deviation of current platform harmfulness $s_t^{(h)}$ from its reference value $r^{(h)}$. At time step $t$, the threshold is computed as
\begin{equation*}
\tau_t =
\operatorname{clip}
\left(
\tau_0
-
\eta
\frac{s_t^{(h)}-r^{(h)}}{r^{(h)}},
\,0,\,1
\right),
\end{equation*}
where $\tau_0$ is the baseline removal threshold and $\eta$ controls how strongly the threshold reacts to the relative deviation of platform harmfulness from its reference. When $s_t^{(h)}>r^{(h)}$, the threshold decreases and removal becomes more aggressive. Conversely, when $s_t^{(h)}<r^{(h)}$, the threshold increases and removal becomes less aggressive. In our experiments, we set $\tau_0=0.7$, corresponding to the threshold used by Fixed Removal, and $\eta=0.2$. For each moderation target $m\in\mathcal{M}_t$, the moderator removes the corresponding post if $v_m\geq\tau_t$ and takes no action otherwise.

\textbf{MPC-based.}
At each time step, moderation targets $m\in\mathcal{M}_t$ are processed sequentially in decreasing order of harmfulness $v_m$. For each target, the moderator evaluates every candidate intervention $a\in\mathcal{A}$ according to its effect on the current platform state, its predicted effect on the next platform state, and an approximation of its persistent future effects over $H$ time steps. Specifically, we define
\begin{equation}
    S_{m,t}(a)
    =
    \mathcal{L}_t(a;m)
    +
    \mathcal{L}_{t+1}(a;m)
    +
    H C_m(a),
    \label{eq:ctms_score}
\end{equation}
where $\mathcal{L}_t(a;m)$ denotes the loss defined in Eq.~\eqref{eq:loss}, evaluated after counterfactually applying intervention $a$ to target $m$, and $\mathcal{L}_{t+1}(a;m)$ denotes the corresponding loss for the predicted platform state at the next time step. 

Targets are evaluated sequentially within each time step. When evaluating target $m$, the current platform state and loss already incorporate the interventions selected for previously processed targets at the same time step. Each candidate action is therefore evaluated conditional on the preceding moderation decisions, allowing their effects to be composed rather than treating targets independently.

To estimate $\mathcal{L}_{t+1}(a;m)$ efficiently, a baseline prediction of the next platform state is first obtained by simulating the entire user population once. For each candidate intervention, this prediction is then updated counterfactually. If $a$ affects the target user's future behavior, only that user is resimulated under the modified behavioral parameters, and their baseline contribution is replaced in the cached aggregate state. Actions that do not affect future user behavior leave the baseline prediction unchanged.

The final term in Eq.~\eqref{eq:ctms_score} approximates effects persisting beyond this one-step prediction. $C_m(a)$ denotes the expected per-time-step capacity cost resulting from applying intervention $a$ to the user associated with target $m$, estimated from its persistent effects on future activity and engagement. Rather than explicitly simulating and jointly optimizing the complete platform trajectory over $H$ time steps, we approximate this longer-term contribution as $H C_m(a)$. For each target $m$, the moderator then selects
\begin{equation*}
    a_m^*
    =
    \arg\min_{a\in\mathcal{A}} S_{m,t}(a).
\end{equation*}

\textbf{PID-based.}
The PID-based moderator uses feedback from the evolving platform to determine how strongly to intervene at each time step. Intuitively, when the platform is far from its desired state, or has remained far from it over time, the moderator requests a larger overall correction. As the platform approaches the desired state, the requested correction decreases. This system-level correction is then distributed across individual moderation targets using the same action-level predictions as the MPC-based moderator.

\begin{figure}[t]
    \centering
    \includegraphics[width=\columnwidth]{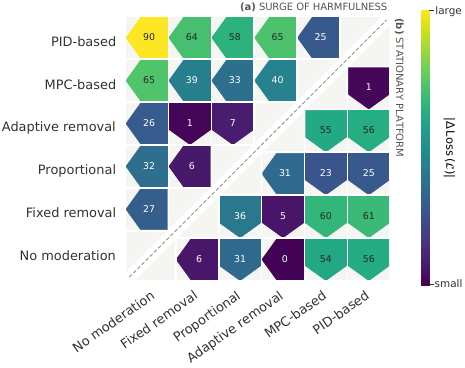}
    \caption{\textbf{Pairwise comparisons between moderators.} Upper triangle: \textbf{(a)} surge of harmfulness. Lower triangle: \textbf{(b)} stationary platform. Values show differences in mean total loss across ten matched simulation runs. Losses include proportionality. For each pairwise comparison, the tip points to the moderator with the lower loss (the better one). All differences are significant at $p<0.01$, based on Holm-corrected two-sided paired $t$-tests within each scenario.}
    \label{fig:pairwise-loss}
\end{figure}

At the beginning of each time step, the moderator measures how far the platform is from its desired state using the loss defined in Eq.~\eqref{eq:loss} as $e_t = \mathcal{L}_{\mathrm{obj}}(\mathbf{s}_t,\mathbf{r})$. The PID controller determines the requested correction from the current error, its accumulation over time, and its recent change:
\begin{equation*}
    \tilde{u}_t
    =
    K_P e_t
    + K_I \sum_{\tau=1}^{t} e_\tau
    + K_D(e_t-e_{t-1}),
\end{equation*}
\begin{equation*}
    u_t=\max\{0,\tilde{u}_t\}.
\end{equation*}
Here, $u_t$ represents the total correction requested at time $t$. The gains $K_P$, $K_I$, and $K_D$ determine how strongly the moderator responds to the current, accumulated, and changing error, respectively. For simplicity, we set $K_P=K_I=K_D=1$ in all experiments, although these parameters can be tuned to obtain different controller responses.

The moderator then distributes this system-level correction across individual targets. Targets $m\in\mathcal{M}_t$ are processed sequentially from highest to lowest harmfulness. For each target, every candidate intervention $a\in\mathcal{A}$ is evaluated using the same score $S_{m,t}(a)$ as for MPC. We define the predicted improvement of an intervention relative to taking no action as
\begin{equation*}
    \Delta_{m,t}(a)
    =
    S_{m,t}(\mathrm{none})-S_{m,t}(a).
\end{equation*}
The moderator selects the intervention whose predicted improvement most closely matches the correction still required. After applying the intervention, its predicted improvement in the controlled platform properties is subtracted from the remaining correction, and the process continues with the next target. Proportionality influences which intervention is selected through $S_{m,t}(a)$, but does not count toward satisfying the system-level correction, which is defined only by the platform objectives.

\begin{figure}[t]
    \centering
    \includegraphics[width=\columnwidth]{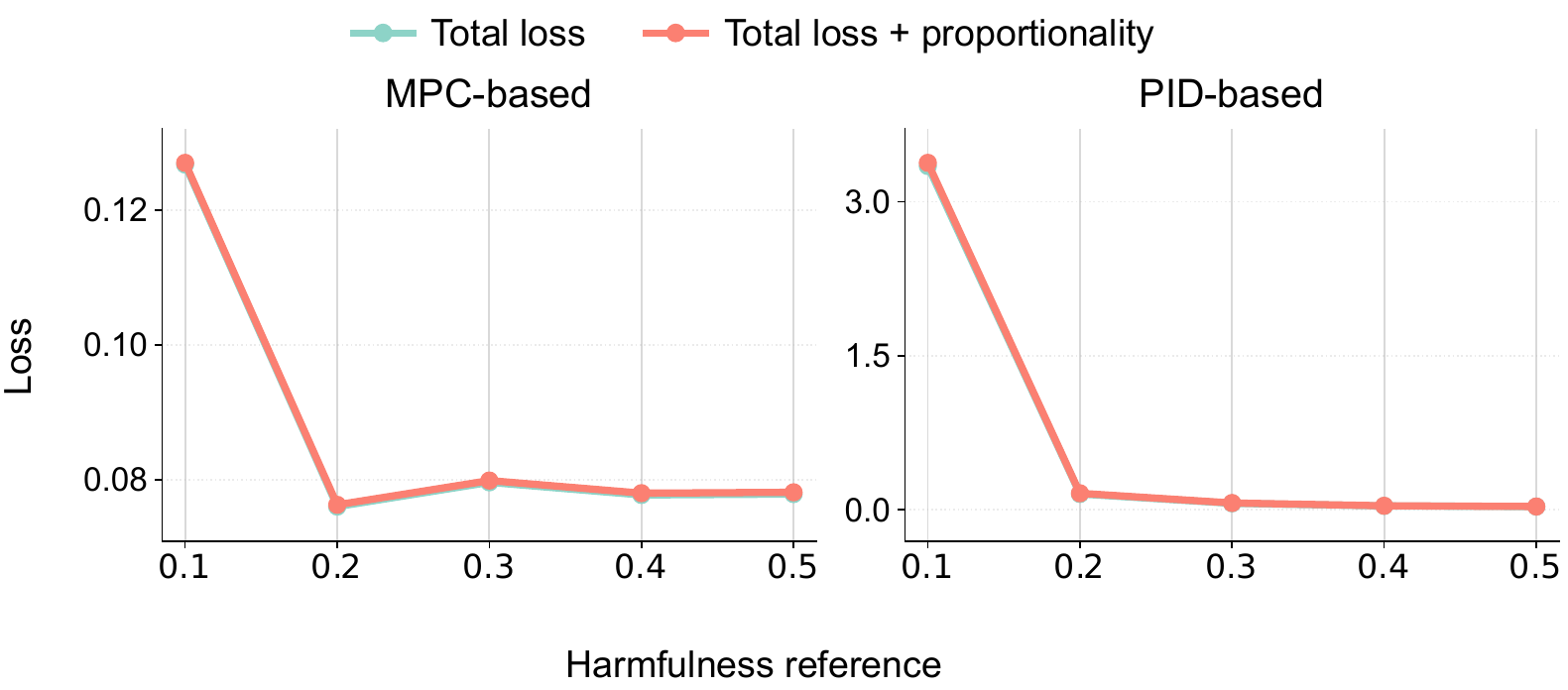}
    \caption{\textbf{Sensitivity analysis: harmfulness reference.} Stationary platform. We report the total loss over harmfulness, activity, and engagement, both with and without the proportionality term, across the different reference values.}
    \label{fig:sensitivity-harmfulness-reference}
\end{figure}

\begin{figure*}[!t]
    \centering
    \includegraphics[width=.9\textwidth]{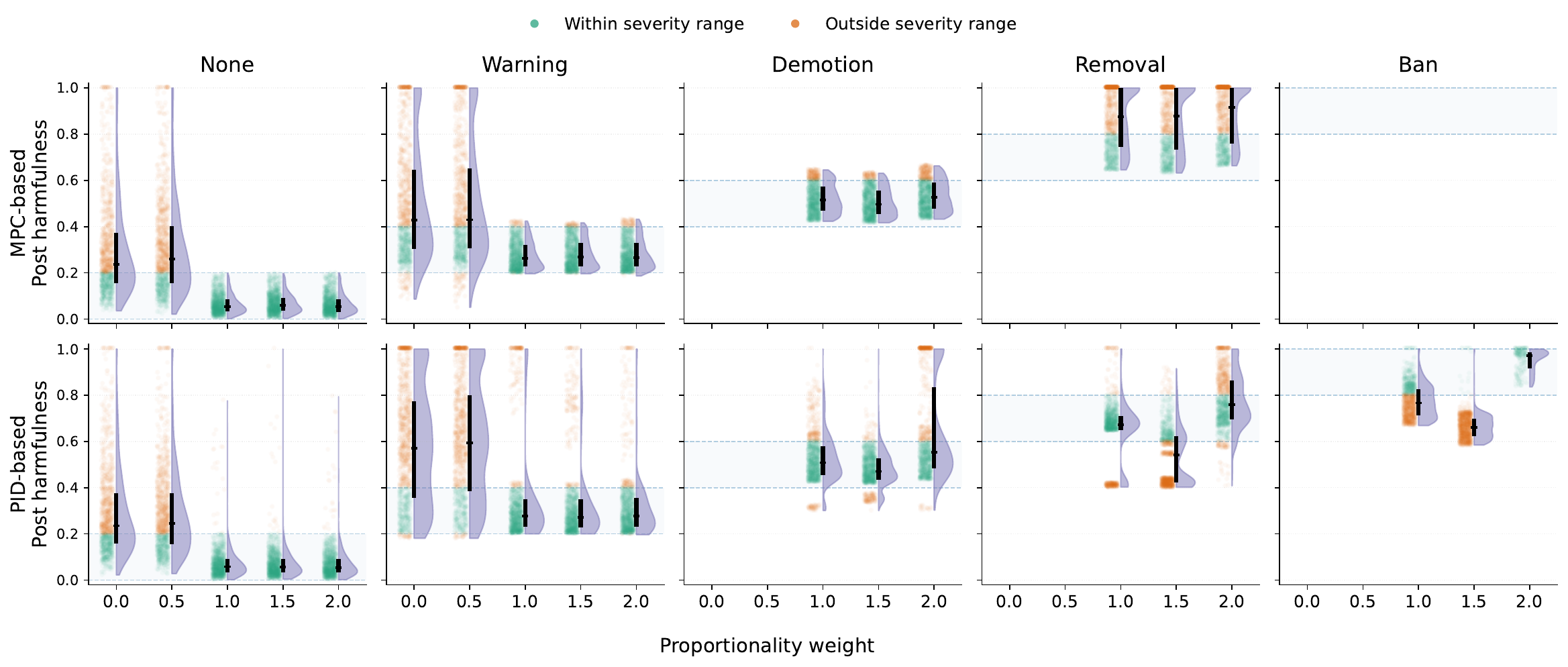}
    \caption{\textbf{Sensitivity analysis: proportionality weight.} Stationary platform. Each column corresponds to a moderation action $a$ and shows the distribution of the harmfulness scores of all posts moderated with $a$, for increasing values of $w_{\mathrm{prop}}$. Green points fall within the intervention's severity range, as per Table~\ref{tab:interventions-effects}, whereas orange points fall outside of it. As $w_{\mathrm{prop}}$ increases, both moderators align intervention severity with post harmfulness, with MPC exhibiting fewer deviations than PID.}
    \label{fig:sensitivity-proportionality}
\end{figure*}

\section{Sensitivity Analyses}
\label{sec:appendix-sensitivity}
We report additional sensitivity analyses and ablation studies under the stationary platform scenario. In particular, we vary key controller parameters, including the lookahead time horizon, proportionality weight, harmfulness weight, and harmfulness reference value, to assess how sensitive the resulting moderation behavior is to these configurations.

Before varying the main controller parameters, we first examine how strongly the moderation strategies differ under the default experimental configuration. Figure~\ref{fig:pairwise-loss} reports pairwise differences in total loss for the stationary and surge scenarios. The comparison shows that the relative separation between strategies is not uniform. Some moderators achieve similar performance, whereas others remain clearly separated, particularly under the surge of harmfulness. All pairwise differences between moderators are statistically significant at $p < 0.01$ based on Holm-corrected two-sided paired $t$-tests.

\subsection{Reference Values}
\label{app:sensitivity-reference-values}
In the main configuration, we set the harmfulness reference to correspond to an average post harmfulness of $\bar{v}=0.3$, such that $r^{(h)}=0.3\,r^{(a)}$. Here, we test alternative reference levels by varying this proportion across the values shown in Figure~\ref{fig:sensitivity-harmfulness-reference}. 
As $\bar{v}$ decreases, moderation becomes increasingly difficult because the harmfulness reference becomes more stringent. At $\bar{v}=0.1$, both moderators incur substantially larger losses, suggesting that this reference is too demanding relative to the other simulation parameters for either moderator to effectively maintain. However, for all tested values $\bar{v}>0.1$, both moderators achieve substantially lower and relatively stable losses, indicating that they can adapt effectively across a range of harmfulness references.

\begin{figure}[t]
    \centering
    \includegraphics[width=\columnwidth]{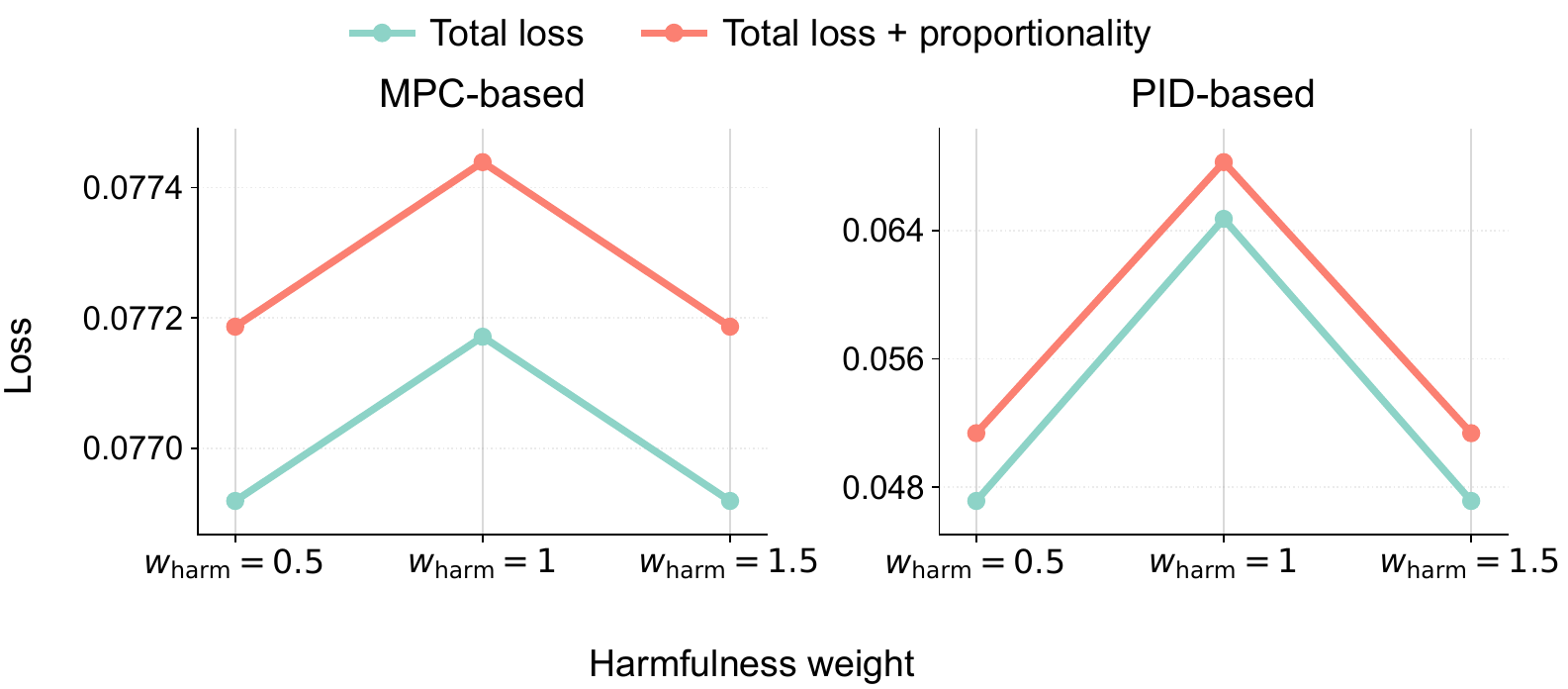}
    \caption{\textbf{Sensitivity analysis: harmfulness weight.} Stationary platform. We report the total loss over harmfulness, activity, and engagement, both with and without the proportionality term, for $w_{\mathrm{harm}} \in \{0.5, 1, 1.5\}$.}
    \label{fig:sensitivity-harmfulness-weight}
\end{figure}

\begin{figure*}[!t]
    \centering
    \includegraphics[width=.9\textwidth]{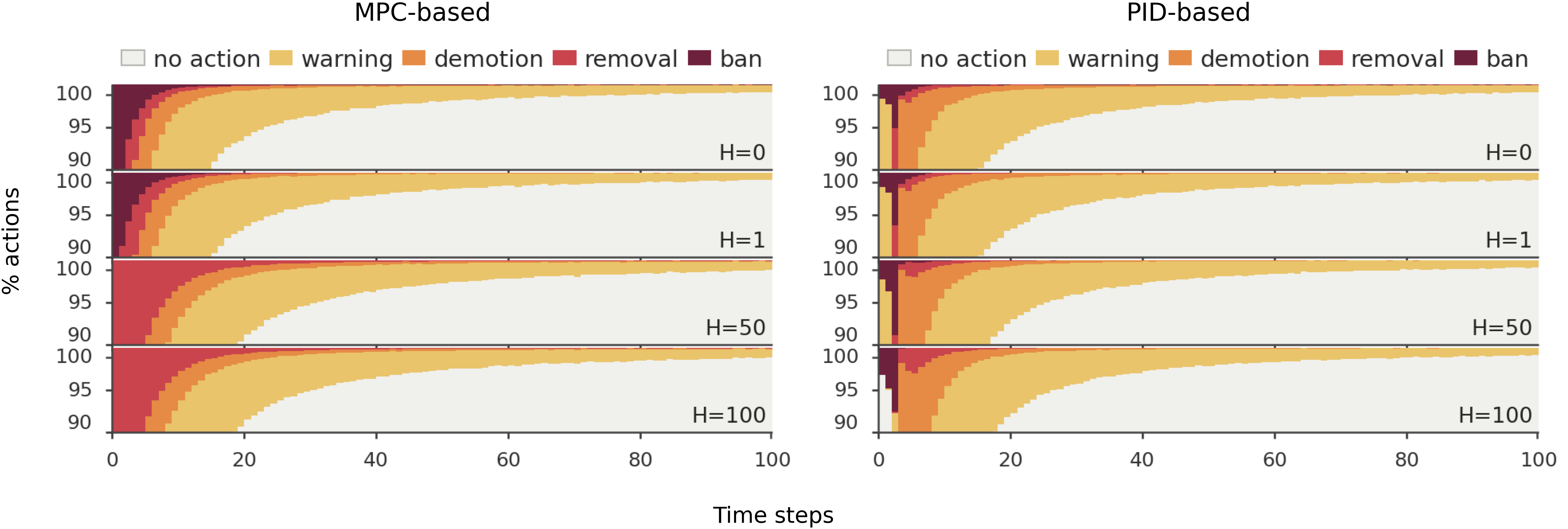}
    \caption{\textbf{Sensitivity analysis: time horizon length.} Stationary platform. Each row shows the proportion of moderation actions over time for $H\in\{0,1,50,100\}$.}
    \label{fig:sensitivity-horizon-actions}
\end{figure*}

\begin{figure}[!t]
    \centering
    \includegraphics[width=\columnwidth]{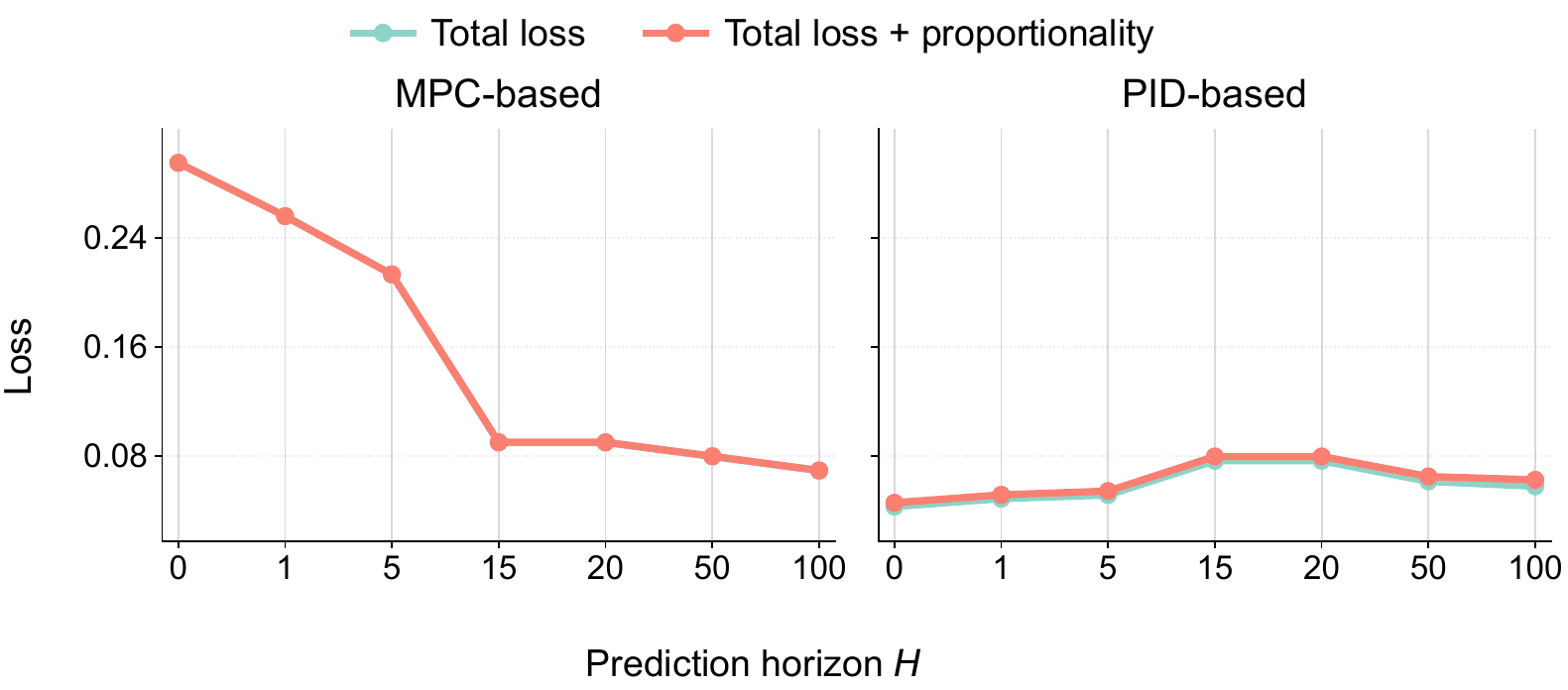}
    \caption{\textbf{Sensitivity analysis: time horizon length.} Stationary platform. We report the total loss over different time horizon lengths, activity, and engagement, both with and without the proportionality term.}
    \label{fig:sensitivity-horizon-loss}
\end{figure}

\subsection{Objective Weights}
\label{app:sensitivity-objective-weights}

\textbf{Proportionality.}
Here, we examine the influence of the proportionality weight on the objective function.
Figure~\ref{fig:sensitivity-proportionality} shows the distribution of post harmfulness for each intervention as $w_{\mathrm{prop}}$ increases. At low proportionality weights, both controllers make extensive use of actions outside their ideal severity ranges, sometimes even choosing not to use more severe moderation actions. As the weight increases, interventions become progressively more aligned with the corresponding harmfulness intervals. This effect is particularly evident for MPC. From $w_{\mathrm{prop}}=1$ and higher, the harmfulness distributions become sharply separated across intervention types, with warning, demotion, and removal largely concentrated within their respective ranges. The controller also shifts low-harmfulness posts toward no intervention rather than applying unnecessarily severe actions.

PID exhibits the same overall tendency, but remains more flexible in its allocation of interventions. Even at higher proportionality weights, some actions are selected outside their ideal ranges, especially for warning, removal, and ban. This pattern is consistent with what reported in Table~\ref{tab:main-results}. At $w_{\mathrm{prop}}=1$, MPC incurs negligible proportionality loss, whereas PID accepts some proportionality loss in exchange for larger gains on the other platform objectives. 

\textbf{Harmfulness.}
Figure~\ref{fig:sensitivity-harmfulness-weight} shows the effect of varying the harmfulness weight $w_{\mathrm{harm}}$. As shown, both moderators incur their highest loss when all platform objectives are weighted equally (i.e., $w_{\mathrm{harm}}=1$, as for the other objectives). This behavior is particularly evident for PID. One possible explanation is that asymmetric weighting provides a clearer prioritization when objectives conflict, whereas equal weighting requires the controllers to balance competing objectives more evenly. Overall, however, performance remains relatively stable across the tested weights, especially for MPC.

\subsection{Time Horizon Length}
\label{app:sensitivity-time-horizon}
\textbf{Prediction horizon.}
Figure~\ref{fig:sensitivity-horizon-actions} shows that the prediction horizon can substantially change which interventions are selected, particularly for MPC. At $H=0$ and $H=1$, bans constitute a substantial share of its early interventions. With longer horizons ($H=50$ and $H=100$), however, bans largely disappear and are replaced by less severe actions, particularly removals. This suggests that a short lookahead is insufficient to capture the persistent negative effects of bans on activity and engagement, and only when these consequences are considered over a longer horizon do bans become comparatively unattractive. PID's intervention composition, by contrast, remains relatively stable across horizons, indicating less sensitivity to the precise value of $H$.

Warnings provide a complementary example. Although they have no immediate platform effect, both moderators still use them at $H=0$ because proportionality favors warnings over no action for posts in the corresponding severity range. Without proportionality, warnings would provide no advantage over no action at $H=0$. With positive horizons, however, their effects on future user behavior also enter action evaluation. Overall, increasing the horizon changes intervention composition most visibly for actions with substantial persistent effects.

Figure~\ref{fig:sensitivity-horizon-loss} further shows that the effect of the horizon on aggregate performance differs between the two moderators. For MPC, loss decreases substantially as the horizon increases, with most of the improvement occurring by $H=15$ and smaller gains thereafter. PID is less sensitive to the horizon. Losses vary across the tested values, but the differences between $H=0$ and $H=1$, $H=50$, and $H=100$ are not statistically significant. Importantly, $H=0$ can be interpreted as an ablation of the predictive component of the PID-based moderator, since intervention selection then relies on PID feedback without accounting for persistent future intervention effects. The absence of a significant performance degradation under this ablation suggests that, in this setting, PID's system-level feedback provides effective temporal adaptation even without the explicit lookahead. In contrast, longer-term prediction plays a more important role for MPC.

\end{document}